\documentclass[12pt]{article}
\usepackage{epsf,amssymb,psfrag}

\newcommand\sect[1]{\setcounter{equation} 0\section{#1}}

\renewcommand\theequation{\thesection\arabic{equation}}   % This is (2.10)
\newcommand{\n}{v}

\newcommand\gdo{\frac{g^2}{4\pi^{D/2}}}

\newcommand\lot[8]{\int_{#1}^{#2} dt_1\int_{#3}^{#4}dt_2
            \int_{#5}^{#6}dt_3\int_{#7}^{#8}dt_4}
\newcommand\coef{\left(\frac{\Gamma(D/2-1)}{4\pi^{D/2}}\right)^2}

\newcommand{\be}{\begin{equation}}
\newcommand{\ee}{\end{equation}}
\newcommand{\ba}{\begin{eqnarray}}
\newcommand{\ea}{\end{eqnarray}}
\newcommand{\baa}{\begin{eqnarray*}}
\newcommand{\eaa}{\end{eqnarray*}}

\newcommand{\ci}[1]{\cite{#1}}
\newcommand{\bi}[1]{\bibitem{#1}}
\newcommand{\lab}[1]{\label{#1}}             %  \mbox{\# {#1}}}
\newcommand{\re}[1]{(\ref{#1})}

\newcommand\e{\mbox{e}}
\newcommand\CO{{\cal O}}
\newcommand\CP{{\cal P}}
\newcommand\CT{{\cal T}}

\newcommand\qqquad{\qquad\quad}
\newcommand\qqqquad{\qquad\qquad}

\newcommand\vev[1]{\langle{#1}\rangle}

\newcommand\bra[1]{|{#1}\rangle}
\newcommand\fracs[2]{\mbox{\small $\frac{#1}{#2}$}}

\def\bom#1{\mbox{\bf{#1}}}
\def\half{\mbox{\small $\frac{1}{2}$}}

\def\frac#1#2{ {{#1} \over {#2} }}

\def\beq{\begin{equation}}
\def\eeq{\end{equation}}
\def\beeq{\begin{eqnarray}}
\def\eeeq{\end{eqnarray}}

\def\fun#1#2{\lower3.6pt\vbox{\baselineskip0pt\lineskip.9pt
  \ialign{$\mathsurround=0pt#1\hfil##\hfil$\crcr#2\crcr\sim\crcr}}}
\def\ie{\hbox{\it i.e.}{ }}      
\def\eg{\hbox{\it e.g.}{ }}      
\def\partder#1{{\partial   \over\partial #1}}
\def \as{\relax\ifmmode\alpha_s\else{$\alpha_s${ }}\fi}
\def \alpi {\frac \as \pi}
\def\ds#1{\ooalign{$\hfil/\hfil$\crcr$#1$}}
\def\MS{\overline{MS}}
\def\bit{\begin{itemize}}
\def\eit{\end{itemize}}
\def\ben{\begin{enumerate}}
\def\een{\end{enumerate}}

   \def\rstyle#1#2#3{#1 (19#3) #2}

\def\np#1#2#3{Nucl.\ Phys.\                   \rstyle{B#1}{#2}{#3}}
\def\pl#1#2#3{Phys.\ Lett.\                   \rstyle{#1B}{#2}{#3}}

\def\prep#1#2#3{Phys.\ Rep.\                  \rstyle{#1}{#2}{#3}}

\def\rmp#1#2#3{Rev.\ Mod.\ Phys.\             \rstyle{#1}{#2}{#3}}

\def\zp#1#2#3{Z.\ Phys.\                   \rstyle{C#1}{#2}{#3}}

\begin{document}
\renewcommand{\thefootnote}{\fnsymbol{footnote}}

\thispagestyle{empty}

\hfill\parbox{35mm}{{\sc UPRF--92--354} \par
                     hep-ph/9210281     \par
                     September, 1992}

\vspace*{15mm}

\begin{center}
{\large \bf Partonic distributions for large $x$ and
\\[2mm]
renormalization of Wilson loop~\footnote{Research supported in part by Ministero
dell'Universt\`a e della Ricerca Scientifica e Tecnologica.} } \vspace{10mm}

{\sc G.~P.~Korchemsky}%
\footnote{On leave from the Laboratory of Theoretical Physics,
          JINR, Dubna, Russia}
\footnote{INFN Fellow}
%
%\medskip
%
and
%
%\medskip
%
{\sc G.~Marchesini}

\bigskip

\medskip

{\em Dipartimento di Fisica, Universit\`a di Parma and \par
INFN, Gruppo Collegato di Parma, I--43100 Parma, Italy} \par
%{\tt e-mail: korchemsky@parma.infn.it; marchesini@parma.infn.it}

\end{center}

\vspace*{20mm}

\begin{abstract}
We discuss the relation between partonic distributions near the phase space
boundary and Wilson loop expectation values calculated along paths partially
lying on the light-cone. Due to additional light-cone singularities,
multiplicative renormalizability for these expectation values is lost.
Nevertheless we establish the renormalization group equation for the light like
Wilson loops and show that it is equivalent to the evolution equation for the
physical distributions. By performing a two-loop calculation we verify these
properties and show that the universal form of the splitting function for large
$x$ originates from the cusp anomalous dimension of Wilson loops.
\end{abstract}

\newpage

\sect{Introduction}

It is well known \ci{IR} that for hard distributions such as deep inelastic
structure functions, fragmentation functions, Drell-Yan pair cross section, jet
production etc.\ near the phase space boundary the perturbative expansion
involves large corrections which needed to be summed~\ci{add2,add3}. This region
is characterized by the presence of two large scales $Q^2$ and $M^2$ with $Q^2
\gg M^2 \gg \Lambda_{\rm QCD}^2$ and at any order in perturbation theory the
leading contributions are given by double logarithmic terms
$(\as^n\ln^{2n}Q^2/M^2).$ The leading contributions can be summed and the result
is given by the exponentiation of the one-loop contribution~\ci{add4}. This
exponentiated form holds also after the summation of the next-to-leading
contributions $(\as^n \ln^{2n-1} Q^2/M^2)$. The exponent is simply modified by
two-loop corrections in which the term $\as^2 \ln^4 Q^2/M^2$ is absent and the
coefficient of $\as^2 \ln^3 Q^2/M^2$ is proportional to the one-loop
beta-function~\ci{add5}. This fact suggests us to apply renormalization group
approach for summing large perturbative contributions to all orders.

The universal properties of hard processes near the phase space boundary
originate from the universality of soft emission, which is responsible of the
double logarithmic contributions. Soft gluon emission from fast quark line can be
treated by eikonal approximation for both incoming and outgoing quarks. In the
eikonal approximation quarks behave as classical charged particles and their
interaction with soft gluons can be described by path ordered Wilson lines along
their classical trajectories~\ci{add6,add7}. The corresponding hard distribution
is then given by the vacuum averaged product of the time-ordered Wilson line,
corresponding to the amplitude, and the anti-time ordered Wilson line,
corresponding to the complex conjugate amplitude. The combination of these Wilson
lines forms a path ordered exponential $W(C)$ along a closed path $C$ which lies
partially on the light-cone and depends on the kinematic of the hard process
\be
\lab{def}
W(C) \equiv \vev{0|\CP\exp\left(ig\oint_{C}dz_\mu A^\mu(z)\right)|0}\,,
\qquad
A_\mu(z)=A_\mu^a(z)\lambda^{a}
\ee
where $\lambda^{a}$ are the gauge group generator. Here the gluon operators
are ordered along the integration path $C$ and not according to time.
Notice that on different parts of the path $C$ the gluon fields are
time or anti-time ordered. This expression is different from the usual
Wilson loop expectation value in which one has ordering along the path for
the colour indices of $\lambda^{a}$ and ordering in time for the gluons fields
$A_\mu^{a}(z)$. In eq.~\re{def} we have ordering both of colour matrices
and gauge fields along the path.

In this paper we study the renormalization properties of $W(C)$ for a path $C$
partially lying on the light-cone and show that renormalization group (RG)
equation for $W(C)$ corresponds to the evolution equation \ci{EQ} for the parton
distribution function or the fragmentation function near the phase space
boundary. In particular, we discuss the relation between the ``cusp anomalous
dimension'' of Wilson loop and the splitting function $P(z)$ for
$z \to 1$. %We explicitly perform the two-loop calculation of $W(C)$ with the
%path $C$ fixed by the kinematics of hard process.

In sect.~2 we establish the relation between $W(C)$ and the distribution function
or fragmentation function near the phase space boundary. In sect.~3 we perform
the two-loop calculation in the Feynman gauge and in the $\rm \MS-$regularization
scheme of $W(C)$ with the path $C$ fixed by the kinematics of hard process. In
sect.~4 we deduce the RG equation for $W(C)$ and show in sect.~5 that this
equation gives rise to the evolution equation for the distribution function and
fragmentation function near the phase space boundary. Sect.~6 contains some
concluding remarks.

\sect{Wilson loop and structure function in the soft limit}

Consider the deep inelastic process for an incoming hadron of energy $E$,
longitudinal momentum $P_\ell$ and mass $M$ probed by a hard photon
with momentum $q.$ In the infinite momentum frame we have $P_+ \gg P_-$,
where the light-cone variables for the hadron momentum
$P_\mu=(P_+,\bom{P}_T,P_-)$ are defined by
$$
P_+=(E+P_\ell)/\sqrt 2,  \qqquad
P_-=(E-P_\ell)/\sqrt 2,  \qqquad
\bom{P}_T=(P_1,P_2)=0 \,.
$$
In this frame $xP_+$ is the ``+'' component of the momentum of the quark
probed by the hard photon, where $x=-{q^2}/{2 (P\cdot q)}$ is the Bjorken
variable. For $x\to1$ the hard photon probes at short distances  the quark
which carries nearly all hadron momentum. In this region the hard process
is characterized by two scales: the virtuality of the photon $Q^2=-q^2$,
which gives the short distance scale, and the large distance scale
$(1-x)Q^2$.

In this Section we show that, after factorization of leading twist contributions,
the partonic distributions for $x\to1$ are given in terms of a generalized Wilson
loop in \re{def}. To show this we first make the leading twist approximation and
then we perform the $x\to1$ limit.

\subsection{Summing collinear gluons}

Due to the factorization theorem \ci{FT} the differential cross-section of this
process is given in terms of the partonic distribution function and the partonic
cross section. The partonic distribution function is a universal distribution
measuring the probability to find in the hadron a parton with $x$ fraction of the
$P_+$ momentum. To leading twist ($Q^2 \gg M^2$) the quark distribution function
has the following representation \ci{Col}
\be
F(x,\mu/M)=\int_{-\infty}^{+\infty} \frac{dy_-}{2\pi}
               \e^{-iP_+y_-x} \vev{P|\, \bar\Psi(y)
               \; \CP\exp\left(ig\int_0^y dz_\mu A^{\mu}(z)\right)
               \; \gamma_+\, \Psi(0)\, |P}
\lab{str}
\,,
\ee
where the vector $y_\mu$ lies on the light-cone with components
$$
y_\mu=(y_+,\bom y_T,y_-)=(0,\bom 0, y_-) \, .
$$
The state $\bra{P}$ describes the hadron with momentum $P$ and mass $M.$
Integration over $y_-$ fixes the value of the ``$+$'' component of the total
momentum of the emitted radiation to be $(1-x)P_+.$ The corresponding transverse
and ``$-$'' components are integrated over by fixing $\bom y_T=0$ and $y_+=0.$
Because of these unrestricted integrations the structure function contains
ultraviolet (UV) divergences which are subtracted at the reference point $\mu.$
The quark field operators $\bar\Psi(y)$ and $\Psi(0)$ are defined in the
Heisenberg representation. The path ordered exponential is evaluated along the
line between points $0$ and $y_\mu$ on the light-cone and ensures the gauge
invariance of the nonlocal composite quark--antiquark operator.

%%%%%%%%%%%%%%%%%%%%%%%%%%%%%%%%%%%%%%%%%%%%%%%%%%%%%%%%%%%
\begin{figure}[t]
\psfrag{p}[cc][cc]{$p$} \psfrag{p1}[cc][cc]{$p'$}
\psfrag{k}[cc][rc]{$k$} \psfrag{q}[cc][rc]{$q$}
\centerline{\epsfxsize6.0cm\epsfbox{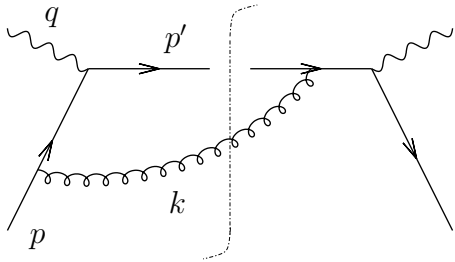}} \caption[]{One-loop Feynman diagram
contributing to the structure function of deep inelastic scattering. The gluon
with momentum $k$ is emitted by the quark $p$ and absorbed by the quark $p'$ in
the final state. We use solid line for quarks, curly lines for gluons,
wavy lines for photons and dot-dashed line for the unitary cut.}%
\end{figure}%
%%%%%%%%%%%%%%%%%%%%%%%%%%%%%%%%%%%%%%%%%%%%%%%%%%%%%%%%%%%%%%%%%%%%

In what follows we will intensively use properties of Wilson lines which will
appear in our analysis. We now briefly describe the physical meaning of Wilson
line entering into the definition \re{str}. Let us start considering the general
form of the structure function of the deep inelastic scattering as matrix element
of two electromagnetic currents:
\be
 \int \frac{d^4 z}{(2\pi)^4}
               \e^{-iqz} \vev{P|j_+(z)j_+(0)|P}
\lab{gen} \,,
\ee
where
$$
j_\mu(z)=\bar\Psi(z)\gamma_\mu\Psi(z)\,.
$$
It is well known that in the leading twist approximation the structure function
is defined by contributions of Feynman diagrams in which only one of the quarks
of the hadron $|{P}\rangle$ participates in the hard scattering \footnote{There
are also contributions initiated by gluons. The corresponding Feynman diagrams
necessary involve real emitted quarks whose contribution is suppressed for $x\to
1$.}. One of these diagrams is shown in fig.~1. In the infinite momentum frame,
in the leading twist limit one can neglect the transverse motion of the quark
inside the hadron and put the velocity of quark equal to the hadron velocity,
\be\lab{vel}
\n_\mu = P_\mu / M\,.
\ee
As we will show in sect.~3, all singularities of the distribution function for
$x\to 1$ are absorbed by Wilson loop depending on the quark velocity $\n$ rather
than momentum.

Consider for example the one-loop contribution of fig.~1 to \re{gen}. In the
infinite momentum frame the incoming quark has momentum $p_\mu$ with $p_- \ll
p_+$ whereas the recoiling momentum $P'$ has component $p'_- \gg p'_+$. The
vertices for the absorption and emission of the real gluon $k$ are given by
$$
\frac{\ds{p}\gamma_\mu(\ds{p}-\ds{k})}{-2pk +i0} \qquad \mbox{and} \qquad
\frac{\ds{p'}\gamma_\mu(\ds{p'}+\ds{k})}{2p'k -i0}
$$
respectively. For vanishing quark mass the two vertices contain the following
singularities: \bit \item $p-$collinear:
     $ k_+ \sim p_+, \quad k_-\sim p_-$
\item $p'-$collinear:
     $ k_+\sim p'_+, \quad k_-\sim  p'_-$
\item soft:
     $\qqqquad k_-\sim k_+ \to 0 \,, $
\eit with $\bom{k}_{\rm T}^2 = 2k_+k_-$. According to the LNK theorem \ci{LNK},
by summing all diagrams for the structure function both the $p'-$collinear and
the soft (or infrared) singularities cancel. We then focus our attention only on
the $p-$collinear singularities. There are also diagrams with quark-photon
vertices ``dressed'' by hard virtual gluons and quarks having momentum
$k_+,k_-,k_{\rm T} \sim Q$. The contribution of hard virtual subprocesses can be
factorized into the partonic cross section.

Consider the vertex for the absorption of gluon $k$ with gauge
potential $a_\mu(k)$ in the momentum representation
\be
g\frac{\ds{p'}\ds{a}(k)(\ds{p'}+\ds{k})}{2p'k -i0} \lab{ver} \,,
\ee
where the gluon field carries longitudinal and transverse polarizations. It can
be easily shown that when the momentum $k$ is collinear to $p$, the components of
$a_\mu(k)$ transverse to $k$ contribute to higher twist and can be neglected.
Thus, only longitudinal polarization survives and the gauge potential becomes
pure gauge field
$$
a_\mu(k)=k_\mu\left(\frac{ya(k)}{yk-i0}\right)\,.
$$
This property allows us to sum to all orders the contribution of
collinear gluons (leading twist approximation) using Ward identities.
Inserting this gauge field in \re{ver} we obtain that the gluon
emission is factorized into
\be
\ds{p'}\cdot g\int \frac{d^4k}{(2\pi)^4}\left(\frac{ya(k)}{yk-i0}\right)
=\ds{p'}\cdot  ig\int d^4x \ J_\mu(x) A^\mu(x) =\ds{p'}\cdot  ig\int_0^\infty
d\tau\ y_\mu A^\mu(y\tau) \,,
\ee
where
$$
J_\mu(x)=y_\mu \int_0^\infty d\tau \ \delta^4(x-y\tau)
$$
is the classical non-abelian eikonal current  of the scattered quark $p'$ in
which we neglect higher twists contributions of order $p'_+/p'_-$. This
factorization property implies that the quark field operator $\bar\Psi(0)$
describing the scattered quark $p'$ in \re{gen} can be written in one-loop as
\be
\lab{one}
\bar\Psi(0)=\bar\Psi_0(0)
\left(1+ig\int_0^\infty dz_\mu A^\mu(z) + \CO(g^2) \right)
\ee
with integration path along $y_-$ axis. Here, $\bar\Psi_0(0)$ is a free quark
operator and $A^\mu(x)$ describes $p-$collinear gluon. Note that this relation is
valid only within the matrix element of currents in \re{gen} and not as an
operator identity.

The generalization of the one loop result \re{one} is based on the following
physical arguments. After the hard scattering the quark with momentum $p'$ starts
to emit gluons and quarks moving in the same direction. All these particles form
$p'-$collinear jet propagating through the cloud of $p-$collinear gluons created
by the incoming quark. Since $p-$collinear gluons are pure gauge fields they
cannot change the state of the scattering quark but only modify its phase. Such a
pure phase factor is given by Wilson line along the light-cone direction $y$
defined in \re{str}. The generalization of \re{one} is then
\be
\lab{col1}
\bar\Psi(0)=\bar\Psi_{p'-{\rm col}}(0)\;\Phi_y[0,\infty;A]
\,,
\ee
where $\Phi_y(z_1,z_2)$ is the path ordered exponential
calculated from point $z_1$ to $z_2$ along the direction $y$
$$
\Phi_y[z_1,z_2;A]\equiv \CP\exp\left(ig\int_{z_1}^{z_2} dz_\mu A^\mu(z)\right)\,.
$$
In \re{col1} the phase $\Phi_y[0,\infty;A]$ describes the interaction of
$p-$collinear gluons with the scattered quark while operator $\bar\Psi_{p'-{\rm
col}}(0)$ describes the jet of $p'-$collinear particles produced by the scattered
quark.

For the quark field operator $\Psi(z)$, which describes absorption of
$p-$collinear gluons by quark with momentum $p'$ in the final state, one obtains
\be\label{col22}
\Psi(z)=\Phi_{-y}[\infty,y;A]\;\Psi_{p'-{\rm col}}(z) \lab{col2} \,,
\ee
with $y=(0,\bom 0,y_-)$ and $y_-=z_-$. After substitution of \re{col1} and
\re{col22} into \re{gen} we get that the full phase of the scattered quark is
given by
\be
\Phi_{-y}[\infty,y;A] \Phi_y[0,\infty;A]=\CP\exp\left(ig\int_0^y dz_\mu
A^\mu(z)\right).
\lab{apr}
\ee
Moreover, the contributions of $p-$collinear and soft particles to the structure
function can be factorized into the distribution function \re{str}. The
contribution of $p'-$collinear jet is factorized as usual into the partonic cross
section~\footnote{As notices in refs.~\ci{add2,add3}, the latter contribution
become large for $x\to 1$. That is why performing the factorization of the
structure function for $x\to 1$ one should treat the $p'-$collinear jet
separately from the partonic cross section.}. Thus, the Wilson line in the
definition of the distribution function \re{str} is the result of the interaction
of scattered quark $p'$ with the gluons collinear to incoming quark $p$. Here, we
have used the following properties of the Wilson lines:
\ba
&\bullet&\  \mbox{hermiticity:}\qqquad
\Phi_y^\dagger[0,\infty;A]=\Phi_{-y}[\infty,0;A]
\nonumber \\
&\bullet&\  \mbox{causality:}\qqqquad
\Phi_y[b,c;A]\Phi_y[a,b;A]=\Phi_y[a,c;A]
\lab{cau}
\\
&\bullet&\  \mbox{unitarity:}\qqqquad
\Phi_y^\dagger[a,b;A]\Phi_y[a,b;A]=1
\nonumber
\ea
Notice, that the approximation \re{apr} is valid in the leading twist
limit for an arbitrary values of the scaling variable $x.$

\subsection{Soft approximation for $x\to 1$}

For $x\to 1$ the total ``+'' momentum of the emitted radiation $(1-x)P_+$
vanishes and the expression \re{str} can be further simplified. In the leading
twist order only $p-$collinear and soft gluons contribute to \re{str}. As already
recalled, due to the LNK theorem $p-$collinear singularities survive in parton
distribution \re{str}. Although soft divergences cancel, their finite
contribution completely determines the asymptotic parton distribution function
for $x\to 1$. To show this we observe that for $x\to 1$, the ``+'' component of
the momentum of any emitted {\it real\/} gluon $q$ is positive and vanishing as
$q_+ \sim (1-x) P_+$. Therefore, $p-$collinear gluons are only virtual since
their momentum $q$ has component $q_+ \sim P_+$. It means that real emitted
gluons are only soft (with momenta $q_+,q_-,q_{\rm T}\sim (1-x)P_+$) while
virtual gluons can be either soft or $p-$collinear.

These conditions imply that before hard scattering by photon probe, the incoming
quark with momentum $p$ decays into a jet of $p-$collinear virtual gluons and
quarks emitting and absorbing soft gluons. The contribution of real and virtual
soft gluons can be factorized from the distribution function as follows. We note
that soft gluons interact with the $p-$collinear quarks and gluons, forming the
$p-$collinear jet, via eikonal vertices. In this approximation the incoming quark
with momentum $p$ behaves as a classical charged particle moving with velocity
$v_\mu$ defined in \re{vel} and all effects of interaction of the $p-$collinear
jet produced by incoming quark with soft gluons are absorbed into phase factors
similar to path-ordered exponentials found in the collinear approximation.
Therefore, in the $x\to 1$ limit one can replace quark field operators in the
definition \re{str} as
\be
\bar\Psi(y)=\bar\Psi_{p-{\rm col}}(y)\Phi_{-v}[y,\infty;A]\,,
\qqqquad \Psi(0)=\Phi_v[\infty,0;A]\Psi_{p-{\rm col}}(0) \lab{sof} \,,
\ee
where the phase factors are evaluated along a trajectory of a massive classical
particle with velocity $v$. Here, the gauge field operators $A_\mu(x)$ describe
soft gluons and the quark fields $\bar\Psi_{p-{\rm col}}(y)$ and $\Psi_{p-{\rm
col}}(0)$ describe the incoming quark in the initial and final states dressed by
$p-$collinear virtual corrections.

We did not consider till now the contribution of quark emissions to the
distribution function. Repeating the previous arguments one finds that for $x\to
1$ {\it real\/} quarks are soft. But it is well known from power counting
\ci{add12} that soft quarks give vanishing contribution to the leading twist
order. Therefore all emitted quarks must be virtual and they contribute either to
$p-$collinear jet or renormalize soft gluon self-interaction vertices. These last
contributions will be explicitly evaluated in subsect.~3.2.

It should be noticed that similar phase factors \re{col1} and \re{col2} describe
interaction of scattered quark with $p-$collinear gluons and incoming quark with
soft gluons, respectively, although the two subprocesses are different. This is
due to the fact that in both cases gluons do not change the quark state: the
collinear gluons are pure gauge fields with vanishing strength; the soft gluons
interact with quarks via eikonal vertices which preserve the momentum and
polarization of the quarks.

Combining the phase factors \re{apr} and \re{sof} from collinear and soft
approximations we obtain the following representation for the distribution
function in the $x\to 1$ limit:
\be
F(x,\mu/M) = H(\mu/M) \int_{-\infty}^{+\infty} P_+\frac{dy_-}{2\pi}
               \e^{iP_+y_-(1-x)} W(C_S)
\lab{main}
\ee
with the function $H(\mu/M)$ taking into account contributions of $p-$collinear
quarks and gluons and
\be
W(C_S)
      =\vev{0|\Phi_{-v}[y,\infty;A]\Phi_y[0,y;A]\Phi_v[\infty,0;A]|0}
      \equiv \vev{0|\CP\exp\left(ig\oint_{C_S}dz_\mu A^\mu(z)\right)|0}
\lab{wl}
\;,
\ee
where the index $S$ indicates that the photon probe has space-like momentum. The
integration path $C_S=\ell_1\cup \ell_2\cup \ell_3$ is shown in fig.~2. The rays
$\ell_1$ and $\ell_3$ correspond to two classical trajectories of a massive
particle going from infinity to $0$ with velocity $v$ and from $y$ to infinity
with velocity $-v,$ respectively. The segment $\ell_2=[0,y]$ lies on the
light-cone. As explained before, since we are dealing with a cross-section rather
than with an amplitude, the gauge fields are ordered along the path rather than
the time. Path- and time-orderings are related as follows
$$
\CP=\CT      \ \mbox{for $\ell_1$}\;, \quad
\CP=\bar \CT \  \mbox{for $\ell_3$}\;, \quad
\CP=\CT      \  \mbox{for $\ell_2$ and $y_->0$}\;, \quad
\CP=\bar \CT \  \mbox{for $\ell_2$ and $y_-<0$}\;, \quad
$$
where $\CT$ and $\bar\CT$ denote time- and anti-time ordering, respectively. This
path has two cusps at the points $0$ and $y$ where the direction of the particle
is changed by the hard photon probe.

%%%%%%%%%%%%%%%%%%%%%%%%%%%%%%%%%%%%%%%%%%%%%%%%%%%%%%%%%%%
\begin{figure}[t]
\psfrag{0}[rc][rc]{$0$} \psfrag{y}[rc][rc]{$y$}
\psfrag{l1}[rc][rc]{$\ell_1$} \psfrag{l2}[rc][rc]{$\ell_2$} \psfrag{l3}[rc][rc]{$\ell_3$}
\centerline{\epsfxsize5.0cm\epsfbox{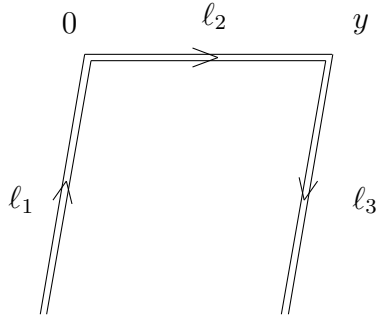}} \caption[]{Integration path
$C_S=\ell_1\cup\ell_2\cup\ell_3$
        for the Wilson loop $W(C_S)$ corresponding to the structure function
        for large $x.$ The ray $\ell_1$ is along the time-like vector
        $\n_\mu$ from $-\infty$ to $0;$ the  segment $\ell_2$ is from
        point $0$ to $y$ along the light-cone; the ray $\ell_3$ is from
        the point $y$ to $-\infty$ along the vector $-\n_\mu$. This path
        has two cusps at points $0$ and $y$ where the quark undergoes hard
        scattering.}%
\end{figure}%
%%%%%%%%%%%%%%%%%%%%%%%%%%%%%%%%%%%%%%%%%%%%%%%%%%%%%%%%%%%%%%%%%%%%

Thus the quark distribution function for $x\to 1$ is given by the product of two
functions: the Fourier transform of the Wilson loop expectation value, which
takes into account all soft emissions and gives the asymptotic distribution
function for $x\to 1,$ and times the function $H(\mu/M)$ getting contribution
from virtual $p-$collinear quarks and gluons. This is the reason why $H(\mu/M)$
does not depend on $P_+(1-x)$ or, equivalently, on $y_-.$ Although $F(x,\mu/M)$
cannot be calculated in perturbation theory we will find its evolution with the
renormalization point $\mu$ using renormalization properties of $W(C_S)$.

\subsection{Fragmentation function in the $x\to 1$ limit}

The Wilson loop representation for the structure function in \re{main}
can be easily generalized for the fragmentation function $D(x,\mu/M)$
in the limit $x\to 1.$ This function gives the probability of finding a
hadron in a quark jet. To the leading twist approximation the fragmentation
function is given by \ci{Col}
\be
D(x,\mu/M)=\int_{-\infty}^{+\infty}\frac{dy_-}{2\pi}
               \e^{-iP_+y_-/x}
             \sum_f \vev{P,f|\bar\Psi(y)\Phi_y[\infty,y;A]|0}
                    \gamma_+\vev{0|\Phi_{-y}[0,\infty;A]\Psi(0)|P,f}
\lab{fra}
\ee
where $P_\mu=(P_+,\bom 0,P_-)$ is momentum of the observed hadron,
$y=(0,\bom 0, y_-)$ is a vector in coordinate space which lies on
the light-cone, and $f$ denotes the associated final state radiation.
In the infinite momentum frame with $P_+\gg P_-$ integration over
$y_-$ fixes the ``+'' component of the total associated radiation
momentum to be $(1/x-1)P_+$. As in Subsection~2.1, the phase factors
in eq.\re{fra} are obtained from the resummation of gluons interacting
with the recoiling quark $P'$ and collinear to the observed hadron $P$.

In the limit $x\to 1$ the associated radiation becomes soft and we can
perform the eikonal approximation as in the previous section. Namely,
the dependence of gauge fields in the quark and anti-quark field operators
in \re{fra} factorizes into two phase factors as in eq.\re{sof}. In
the soft ($x\to 1$) limit we obtain the following representation for
the fragmentation function
$$
D(x,\mu/M) = H(\mu/M)\int_{-\infty}^{+\infty}P_+ \frac{dy_-}{2\pi}
               \e^{iP_+y_-(1-1/x)} W(C_T)
\,.
$$
The generalized vacuum averaged Wilson loop operator $W(C_T)$ is given by
\baa
W(C_T)
       &=&\sum_f\vev{f|\Phi_v[y,\infty;A]\Phi_y[\infty,y;A]|0\rangle
               \langle 0|\Phi_{-y}[0,\infty;A]\Phi_{-v}[\infty,0;A]|f}
\\
   &=&\vev{0|\Phi_{v}[y,\infty;A]\Phi_y[0,y;A]\Phi_{-v}[\infty,0;A]|0}
\\
    &\equiv& \vev{0|\CP\exp\left(ig\oint_{C_T}dz_\mu A^\mu(z)\right)|0}
\,,
\eaa
where we have used completeness condition for the associated radiation
$1=\sum_f |f\rangle\langle f|$ and causality \re{cau} for the phase factors.
The index $T$ indicates that the photon probe has time-like momentum.

The integration path $C_T=\ell_1\cup \ell_2\cup \ell_3$ is shown in fig.~3
and three parts $\ell_i$ have meanings similar to that described before
for the space-like process. In this case path- and time-orderings are
related as follows
$$
\CP= \CT    \    \mbox{for $\ell_1$}\;,
\quad
\CP= \bar \CT        \   \mbox{for $\ell_3$}\;,
\quad
\CP=\CT         \    \mbox{for $\ell_2$ and $y_->0$}\;,
\quad
\CP=\bar \CT    \   \mbox{for $\ell_2$ and $y_-<0$}
$$
This path has two cusps at the points $0$ and $y$ where the direction of
the particle is changed by the hard photon probe.

%%%%%%%%%%%%%%%%%%%%%%%%%%%%%%%%%%%%%%%%%%%%%%%%%%%%%%%%%%%
\begin{figure}[t]
\psfrag{0}[rc][rc]{$0$} \psfrag{y}[rc][rc]{$y$}
\psfrag{l1}[rc][rc]{$\ell_1$} \psfrag{l2}[rc][rc]{$\ell_2$} \psfrag{l3}[rc][rc]{$\ell_3$}
\centerline{\epsfxsize5.0cm\epsfbox{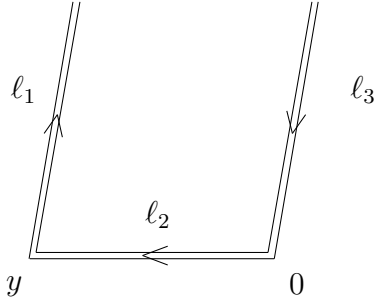}} \caption[]{Integration path $C_T$
corresponding to the fragmentation function
        for large $x.$}%
\end{figure}%
%%%%%%%%%%%%%%%%%%%%%%%%%%%%%%%%%%%%%%%%%%%%%%%%%%%%%%%%%%%%%%%%%%%%

\subsection{One-loop calculation}

We perform the one-loop calculation of $W(C_S)$ in order to explain
in this formulation the nature and origin of double logarithms and to
discuss the analytical properties of $W(C_S)$ in the $y_-$ variable.

We parameterize the integration path $C_S=\ell_1\cup \ell_2\cup \ell_3
=\{z_\mu(t); t\in (-\infty,+\infty)\}$ as follows
$$
z_\mu(t)=\left\{\begin{array}{ll}
                \n_\mu t,& \qquad -\infty < t < 0 \\
                 y_\mu t,&  \qquad 0 < t < 1       \\
                 y_\mu-{\n}_\mu(t-1),&  \qquad 1 < t < \infty
               \end{array}
         \right.,
\qqqquad \n_\mu \equiv \frac {P_\mu}{M}.
$$
The relevant diagrams are given in fig.~4 plus the symmetric ones. We perform the
calculation in the Feynman gauge and in coordinate representation by using the
dimensional regularization. As we shall see in the sect.~3 this representation is
very convenient for the calculation of $W(C_S)$ to two loops. Feynman rules in
coordinate representation are summarized in the Appendix~A.

%%%%%%%%%%%%%%%%%%%%%%%%%%%%%%%%%%%%%%%%%%%%%%%%%%%%%%%%%%%
\begin{figure}[t]
\psfrag{(a)}[cc][cc]{(a)} \psfrag{(b)}[cc][cc]{(b)}
\psfrag{(c)}[cc][rc]{(c)}
\centerline{\epsfxsize14.0cm\epsfbox{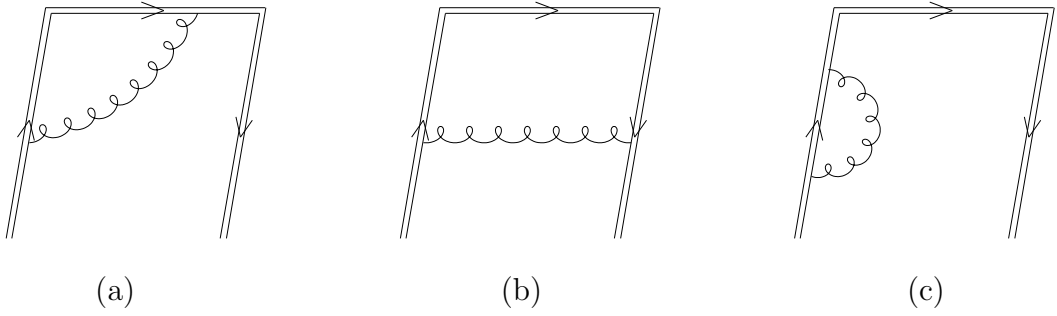}} \caption[]{ One-loop diagrams
contributing to $W(C_S).$ Here, the double line represents the integration path
in the Minkowski space as in fig.~2. The Feynman rules for these diagrams are
given in the Appendix.
}%
\label{fig1}%
\end{figure}%
%%%%%%%%%%%%%%%%%%%%%%%%%%%%%%%%%%%%%%%%%%%%%%%%%%%%%%%%%%%%%%%%%%%%

%%%%%%%%%%%%%%%%%%%%%%%%%%%%%%%%%%%%%%%%%%%%%%%%%%%%%%%%%%%
\begin{figure}[t]
\psfrag{0}[cc][cc]{$0$} \psfrag{y}[cc][cc]{$y$}
\centerline{\epsfxsize14.0cm\epsfbox{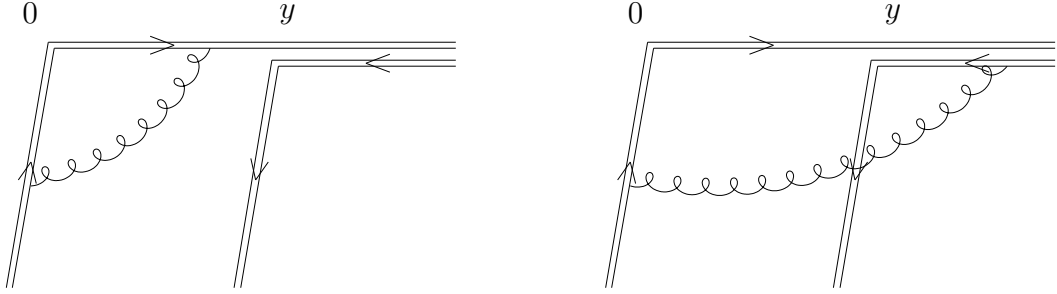}} \caption[]{ Diagrams corresponding
to fig.~4a for virtual and real gluon
        (see \re{l1a}). Due to a partial cancellation of Wilson lines,
        the sum of these two diagram gives the single contribution of
        fig.~4a. Similar cancellations hold for all diagrams.
}%
%\label{fig1}%
\end{figure}%
%%%%%%%%%%%%%%%%%%%%%%%%%%%%%%%%%%%%%%%%%%%%%%%%%%%%%%%%%%%%%%%%%%%%

The path from point $0$ to $y$ consists actually of the product of paths from $0$
to $\infty$ and from $\infty$ to $y$ as shown in fig.~5 for the diagram of
fig.~4a. The gluon is either real (fig.~5a) or virtual (fig.~5b). In both cases
the gluon is emitted at point $z_1=\n t_1$ with $t_1 < 0.$ For real gluon we have
$z_2=yt_2$ with $1<t_2<\infty,$ while for virtual gluon we have $z_2=yt_2$ with
$0<t_2<\infty.$ The two contributions are given by
\be
W^{(1)}_a=(ig)^2C_F \n_\mu y_\nu \int_{-\infty}^0 dt_1 \left\{\int_0^\infty dt_2\
D^{\mu\nu}(z_2-z_1)+\int_\infty^1 dt_2\ D^{\mu\nu}_+(z_2-z_1) \right\} \lab{l1a}
\ee
where $C_F$ is the quadratic Casimir operator in the quark representation. For
$y_-$ positive the vector $z=(z_2-z_1)$ is time-like ($z^2>0$ and $z_0>0$) so
that cutted and full propagators coincide:
$D^{\mu\nu}_+(z)=D^{\mu\nu}(z)=-g^{\mu\nu} D(z).$ Then, the sum of the two
diagrams of fig.~5 is
\be
W^{(1)}_a=g^2 C_F (\n y) \int_{-\infty}^0 dt_1 \int_0^1 dt_2\ D(yt_2-nt_1)
\lab{sum}
\ee
The same result is obtained for $y_-$ negative by deforming the integration path
$\ell_1$. To show this observe first that the singularities of the integrand in
the $t_1-$complex plane lie in the lower half plane. Therefore we can deform the
integration path from $-\infty <t_1 < 0$ to $0 < t_1 < \infty.$ After this
deformation the vector $yt_2-\n t_1$ becomes time-like and we have the same
situation as for $y_-$ positive.

Both the real and virtual contributions in \re{l1a} have infrared (IR)
and UV singularities. The real and virtual IR singularities, coming from
$t_2\to \infty$, cancel in \re{sum}, giving a bonded range for $t_2$.
This cancellation between the two paths from point $y$ to $\infty$ is
general since it is the result of causality of Wilson lines in \re{cau}.
For this reason in fig.~4 we neglect these parts of the integration paths.

Integral in eq.\re{sum} is ultraviolet divergent and we use dimensional
regularization. Using the $D-$dimensional propagator given in the
Appendix we obtain
$$
W_a^{(1)}  = \gdo \mu^{4-D}\Gamma(D/2-1)C_F(\n y)\int_{-\infty}^0 dt_1\int_0^1
dt_2 \left(-t_1^2+2(\n y)t_1t_2+i0\right)^{1-D/2}
$$
which is divergent for $D=4.$ To see the origin of these divergences we change
the integration variable to $t_1=-2(\n y)t_2t_1'$ and obtain
$$
W_a^{(1)} =-\frac{g^2}{8\pi^{D/2}}C_F\Gamma(D/2-1)[2i\mu(\n\cdot y-i0)]^{4-D}
\int_0^\infty dt_1' (t_1'(1+t_1'))^{1-D/2} \int_0^1 dt_2 \,t_2^{3-D}
$$
For $D\to 4$ we find two  singularities: for $t_2\to 0$ and $t_1' \to 0.$
In the first case we have $z_2-z_1\to 0$, which corresponds to the
integration near the cusp at point $0$. This singularity is called
``cusp singularity'' \ci{away}. For $t_1'\to 0$ we have $z_1\to 0$
and $(z_2-z_1)^2 \to z_2^2=0$ which corresponds to the light-cone
(collinear) singularity. The final unrenormalized result for this diagram
is
\be
W_a^{(1)} =-\gdo C_F[2i\mu(\n\cdot y-i0)]^{4-D}
\frac{\Gamma(3-D/2)\Gamma(D-3)}{(4-D)^2} \lab{res:a}
\ee
For the diagram of fig.~4b we obtain
\baa
W_b^{(1)} &=& (ig)^2 C_F \int_{-\infty}^0 dt_1 \int_0^\infty dt_2\,
D_+(y-\n(t_2+t_1))
 \\
   &=&-\gdo \mu^{4-D} \Gamma(D/2-1)C_F\int_{-\infty}^0 dt_1
   \int_0^\infty dt_2
\\ &\times &
   \left[2(y_- -\n_-(t_2+t_1)-i0)(\n_+(t_2+t_1)+i0)\right]^{1-D/2}
\eaa
This integral for $D\to 4$ has an infrared singularity for $z_2-z_1
\to\infty.$ Taking $D > 4$ one gets
\be
W_b^{(1)} =\gdo C_F[2i\mu(\n\cdot
y-i0)]^{4-D}\frac{\Gamma(3-D/2)\Gamma(D-3)}{4-D} \lab{res:b}
\ee
For the diagram of fig.~4c we have
\ba
W_c^{(1)} &=&\gdo\mu^{4-D}\Gamma(D/2-1)C_F \int_{-\infty}^0
dt_1\int_{t_1}^0dt_2\,(-(t_1-t_2)^2+i0)^{1-D/2}
\nonumber
\\
   &=&-\gdo C_F(\mu^2\n^2)^{2-D/2} i^{4-D}\frac{\Gamma(D/2-1)}{3-D}
     \int_{-\infty}^0 dt_1\, t_1^{3-D}
\lab{res:c}
\,.
\ea
In this case we have an IR singularity for $D < 4$ as $t_1-t_2\to\infty$ and an
UV divergence for $D > 4$  as $t_1\to 0$ and $t_1-t_2\to 0.$ The IR divergence of
$W_c^{(1)} $ is cancelled by the IR divergence of $W_b^{(1)} .$ Thus, the sum of
the diagrams $W_b^{(1)} +W_c^{(1)} $ contains only UV divergence. This IR
cancellation does not depend on the scheme we use to regularize IR divergences,
\eg by putting a cutoff in the $t_i-$integration or by giving a fictitious mass
to the gluon.

Notice, that the IR and UV poles in $W_c^{(1)}$ have opposite coefficients,
thus one can formally set $W_c^{(1)} =0.$ In this case however the pole
of $W_b^{(1)} $ for $D=4$ has to be interpreted as an UV singularity.

Summing eqs.\re{res:a}, \re{res:b} and the symmetric contributions
we obtain the one-loop expression for the unrenormalized Wilson loop
\re{wl}
\be
W^{(1)}=\gdo C_F[2i\mu(\n \cdot y-i0)]^{4-D}\Gamma(3-D/2)\Gamma(D-3)
\left(-\frac2{(4-D)^2}+\frac1{4-D}\right). \lab{res:un}
\ee
By subtracting the poles in the $\rm \MS-$scheme we obtain
\be
W^{(1)}_{ren.}=\alpi C_F\left(-L^2 + L - \frac5{24}\pi^2\right), \qquad
L=\ln(i(\rho - i0)) + \gamma_{\rm E} \lab{res:re}
\ee
where
\be
\rho=(\n y)\mu =(Py)\frac\mu{M}, \qqqquad y=(0,\bom 0,y_-)\,. \lab{res:re'}
\ee
{}From eqs.\re{res:un} and \re{res:re} we can directly see that $W(C_S)$ to one
loop depends only on the variable $\rho$ and the possible singularities are in
the upper half plane
\be
W(C_S)=W(\rho-i0)\,.
\lab{ana}
\ee
This is due to the fact that $\rho$ is the only scalar dimensionless variable
formed by $\n$, $y$ and $\mu$. The ``$-i0$'' prescription comes from the position
of the pole in the free gluon propagator in the coordinate representation.
Moreover, expression \re{res:re} implies that under complex conjugation
$$
W(\rho-i0) = \left(W(-\rho-i0 \right)^*\,.
$$
It is this property that ensures the reality condition of the distribution
function \re{main}.

\sect{Two-loop calculation}

In this section we perform the two-loop calculation of $W(C_S)$.
In order to simplify the calculation and reduce the number of diagrams
to compute we use the nonabelian exponentiation theorem \ci{ET}.
According to this theorem we can write
$$
W \equiv 1+\sum_{n=1}^\infty \left(\alpi\right)^n W^{(n)}
      =  \exp \sum_{n=1}^\infty \left(\alpi\right)^n w^{(n)}
$$
where $w^{(n)}$ is given by the contributions of $W^{(n)}$ with the
``maximal nonabelian color and fermion factors'' to the $n-$th order
of perturbation theory. At one loop we have only the color factor $C_F.$
At two-loops the maximal nonabelian color factor is $C_AC_F$ and the
fermionic factor is $C_FN_f.$ Here, $C_A$ is the quadratic Casimir
operator in the gluon representation and $N_f$ the number of light
quarks. From this theorem we have $W^{(1)}=w^{(1)}$ and
$$
W^{(2)}=\half(w^{(1)})^2 + w^{(2)}
$$
and the contributions with the abelian color factor $C_F^2$ are
contained only in the first term.

In fig.~6 we list all nonvanishing the two-loop diagrams which contain
color factors $C_FC_A$ and $C_FN_f.$ As for the one loop case, the paths
in these diagrams do not contain the two rays from $y$ to $\infty$ because
their contributions cancel due to causality of Wilson lines in \re{cau}.

The color factor of abelian-like diagrams of figs.~6.1-6.6 is
$C_F(C_F-\half C_A).$ These diagrams contribute to $w^{(2)}$ with
the color factor $-\half C_FC_A.$ Diagrams with gluon self-energy of
figs.~6.7 and 6.8 will contain contributions with both color factors
$C_FC_A$ and $C_FN_f.$ The latter one comes from the quark loop
contribution. Finally, diagrams of figs.~6.9-6.11 are of nonabelian
nature involving three-gluon coupling and their color factor is
proportional to $C_FC_A.$ We have omitted diagrams with abelian color
factor $C_F^2$, diagrams vanishing due to the antisymmetry of three-gluon
vertex, diagrams proportional to $y^2=0$ and self-energy like diagrams
obtained by iterating the one-loop diagram in fig.~4c. The reason for
neglecting of the latter type of diagrams is the following. As in the
one-loop case the self-energy diagrams have both IR and UV poles which
cancel each other. On the other hand in the sum of all two loop diagrams
the IR singularities cancel completely. Therefore, as we did for one-loop
case, we neglect the self-energy diagrams and interpret the IR poles in
the diagrams of fig.~6 as UV singularities.

For the diagrams of figs.~6.2--6.5 and figs.~6.8 and 6.10 we have to
consider contributions involving cutted propagators. As we have observed
before, there is no difference between real and virtual gluon propagating
in the time-like direction ($D(z)=D_+(z)$ for time-like $z$). This gives
rise to a partial cancellation between real and virtual contributions.
This cancellation has been already used to simplify the analysis in the
one-loop case \re{res:c} and we will further exploit it in this section.
It turns out that to compute singular contributions for $D=4$ we can
replace the cutted propagators in fig.~6 by full ones. This is due to
the fact that one can always deform the integration path in such a way
that gluons propagate in the time-like direction.

%%%%%%%%%%%%%%%%%%%%%%%%%%%%%%%%%%%%%%%%%%%%%%%%%%%%%%%%%%%
\begin{figure}[t]
\psfrag{1}[cc][cc]{$\scriptstyle 1$} \psfrag{2}[cc][cc]{$\scriptstyle 2$}
\psfrag{3}[cc][cc]{$\scriptstyle 3$} \psfrag{4}[cc][cc]{$\scriptstyle 4$}
\psfrag{(1)}[cc][cc]{$(1)$} \psfrag{(2)}[cc][cc]{$(2)$}
\psfrag{(3)}[cc][cc]{$(3)$} \psfrag{(4)}[cc][cc]{$(4)$}
\psfrag{(5)}[cc][cc]{$(5)$} \psfrag{(7)}[cc][cc]{$(7)$}
\psfrag{(6)}[cc][cc]{$(6)$} \psfrag{(8)}[cc][cc]{$(8)$}
\psfrag{(9)}[cc][cc]{$(9)$} \psfrag{(10)}[cc][cc]{$(10)$}
\psfrag{(11)}[cc][cc]{$(11)$}
\centerline{\epsfxsize16.0cm\epsfbox{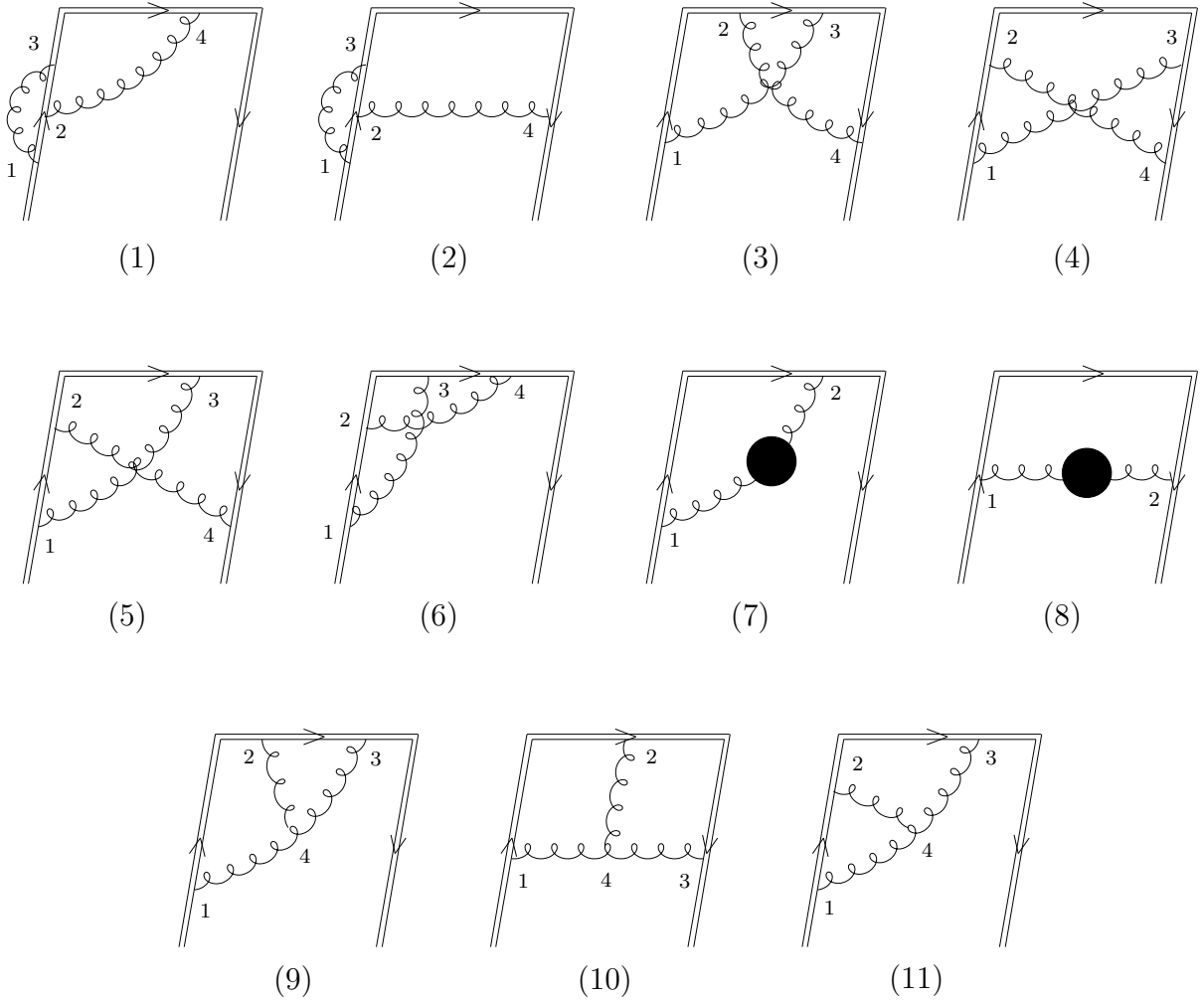}} \caption[]{Nonvanishing two-loop
diagrams containing the ``maximally
        nonabelian'' color $C_AC_F$ and the fermionic $C_FN_f$ factors.
        Due to the exponentiation theorem, these are the only diagram we
        need to evaluate to compute the two-loop contribution of $W(C_S)$.
        The blob denotes the sum of gluon, quark and ghost loops.}%
\end{figure}%
%%%%%%%%%%%%%%%%%%%%%%%%%%%%%%%%%%%%%%%%%%%%%%%%%%%%%%%%%%%%%%%%%%%%

\subsection{Abelian like diagrams}

The general form of the contribution of the abelian like diagrams of
figs.~6.1--6.6 is ($a=1,\ldots,6$)
\be
W_a=\left(\gdo\right)^2 C_F(C_F-\half C_A) [2i(\rho -i0)]^{8-2D} I_a(D) \lab{abe}
\ee
where $I_a(D)$ is a function of $D$ and the index $a$ refers to the
diagrams in fig.~6 and $\rho$ is given in \re{ana}.

Diagram of fig.~6.1 gives
$$
W_1=g^4 C_F(C_F-\half C_A) (\n y)
\lot{-\infty}0{t_1}0{t_2}001 \, D(z_3-z_1)D(z_4-z_2)
$$
with
$$
D(z_3-z_1)D(z_4-z_2)=
\coef
\left[ (t_1-t_3)^2t_2(t_2-\frac{y_-}{\n_-}t_4)\right]^{1-D/2}
$$
where $z_i=\n t_i,$ $i=1,2,3$ and $z_4=yt_4.$
This diagram contain UV divergence for $z_1-z_3\to 0$ corresponding to
vertex renormalization. The rest of the diagram has the same divergences
as one loop diagram of fig.~4a. Namely, it has a cusp singularity for
$z_4-z_2\to 0$ and light-cone collinear singularity for $z_2\to 0.$
Performing the integration we obtain
$$
I_1=-\frac{\Gamma(D/2-1)\Gamma(7-3D/2)\Gamma(2D-7)}{6(3-D)(4-D)^3}
$$
Diagram of fig.~6.2 gives
$$
W_2=-g^4 C_F(C_F-\half C_A)
\lot{-\infty}0{t_1}0{t_2}00\infty \,D(z_3-z_1)D_+(z_4-z_2)
$$
with
$$
D(z_3-z_1)D_+(z_4-z_2)=\coef
\left[(t_1-t_3)^{2}(t_2+t_4+i0)
(t_2+t_4-\frac{y_-}{\n_-}+i0)\right]^{1-D/2}
$$
where $z_i=\n t_i,$ $i=1,2,3$ and $z_4=y-\n t_4.$ The analysis of
singularities is similar to the previous case. We have a UV singularity
for $z_1-z_3\to 0$ and a IR pole originated from one-loop diagram of
fig.~4b. The result of the integration is
$$
I_2=\frac{\Gamma(D/2-1)\Gamma(7-3D/2)\Gamma(2D-7)}{2(3-D)(4-D)^2(5-D)}
$$
Diagram of fig.~6.3 gives
$$
W_3=-g^4 C_F(C_F-\half C_A) (\n y)^2
\lot{-\infty}001{t_2}10\infty \,D(z_3-z_1)D_+(z_4-z_2)
\,,
$$
where $z_1=\n t_1,$ $z_2=yt_2,$ $z_3=yt_3$ and $z_4=y-\n t_4.$ Since
all singularity in $t_4-$plane lie at the lower half-plane we deform
the $t_4-$integration path from $[0,\infty)$ to $(-\infty,0].$ After
this transformation $z_4$ is replaced by $z_4'=y+\n t_4$ and $z_4'-z_2$
becomes time-like vector for which cutted and full propagators are the
same. We can than replace
$$
D(z_3-z_1)D_+(z_4-z_2) \Rightarrow D(z_3-z_1)D_+(z_4'-z_2)
= D(z_3-z_1)D(z_4'-z_2)
$$
which gives
$$
D(z_3-z_1)D(z_4'-z_2) = \coef \left[t_1 t_2(t_1-\frac{y_-}{\n_-}t_3)
(t_4+\frac{y_-}{\n_-}(1-t_2))\right]^{1-D/2} \,.
$$
This diagram has two light-cone collinear singularities for $z_1\to 0$ and
$z_4\to y.$ Evaluating the integral we obtain
$$
I_3=\frac{\Gamma^2(3-D/2)\Gamma^2(D-3)}{(4-D)^4}
\left(1-\frac{\Gamma^2(5-D)}{\Gamma(9-2D)}\right)
\,.
$$
The diagram of fig.~6.4 gives
$$
W_4= g^4 C_F(C_F-\half C_A)
\lot{-\infty}0{t_1}00\infty{t_3}\infty \,D_+(z_3-z_1)D_+(z_4-z_2)
$$
where $z_i=\n t_i$ for $i=1,2$ and $z_j=y-\n t_j$ for $j=3,4.$ As in the
previous diagram we can deform $z_3$ and $z_4$ integration paths by
replacing $z_j=y-\n t_j$ with $z_j'=y+\n t_j$ for $j=3,4$ corresponding to
the replacement
$$
D_+(z_3-z_1)D_+(z_4-z_2) \Rightarrow D_+(z_3'-z_1)D_+(z_4'-z_2)
= D(z_3'-z_1)D(z_4'-z_2)
$$
and we have
$$
D(z_3'-z_1)D(z_4'-z_2)=\coef\left[ (t_4-t_2+\frac{y_-}{\n_-})(t_4-t_2)
(t_3-t_1+\frac{y_-}{\n_-})(t_3-t_1) \right]^{1-D/2}.
$$
%\baa
%D(z_3'-z_1)D(z_4'-z_2) &=&
%\\
%& & \hspace*{-40mm}
%\coef\left[
%(t_4-t_2+\frac{y_-}{n_-})(t_4-t_2)
%(t_3-t_1+\frac{y_-}{n_-})(t_3-t_1)
%\right]^{1-D/2}
%\,.
%\eaa
This diagram has a single IR pole for $z_4-z_2\to\infty$ and
$z_3-z_1\to\infty$ simultaneously. Evaluating the integral we obtain
$$
I_4=-\frac1{4-D}\frac{\Gamma(2D-7)}{2\Gamma^2(D/2-1)}+\CO((4-D)^0)
\,.
$$
Diagram of fig.~6.5 gives
$$
W_5=-g^4 C_F(C_F-\half C_A)
\lot{-\infty}0{t_1}0010\infty \,D(z_3-z_1)D_+(z_4-z_2)
\,,
$$
where $z_i=\n t_i$ for $i=1,2,$ $z_3=yt_3$ and $z_4=y-\n t_4.$
Deforming the $z_4$ integration path we replace $z_4$ with $z_4'=y+\n
t_4$
$$
D(z_3-z_1)D_+(z_4-z_2) \Rightarrow D(z_3-z_1)D_+(z_4'-z_2)
= D(z_3-z_1)D(z_4'-z_2)
$$
and we have
$$
D(z_3-z_1)D(z_4'-z_2)=\coef
\left[t_1(t_1-\frac{y_-}{\n_-}t_3)(t_4-t_2+\frac{y_-}{\n_-})(t_4-t_2)
\right]^{1-D/2} \,.
$$
It turns out that this diagram has no singularities for $D=4.$
To confirm this we evaluate the integral and obtain
\baa
I_5&=&-\Gamma^2(D/2-1)\left\{-\frac{\pi^2}3 \right.
    +\frac1{(4-D)^3(3-D)}
     \left(
       \frac{3\Gamma(3-D/2)\Gamma(2D-7)}{\Gamma(3D/2-5)}
\right.
\\
& &- \left.\left.
       \frac{\Gamma(5-D)\Gamma(2D-7)}{\Gamma(D-3)}
      -\frac{2\Gamma(3-D/2)\Gamma(D-3)}{\Gamma(D/2-1)}
    \right)
\right\} = \CO((4-D)^0)
\,.
\eaa
Diagram of fig.~6.6 gives
$$
W_6=g^4 C_F(C_F-\half C_A)  (\n y)^2 \lot{-\infty}0{t_1}001{t_3}1 \, D(z_3-z_1)
D(z_4-z_2)
$$
where $z_i=\n t_i$ for $i=1,2,$ $z_j=yt_j$ for $j=3,4,$
and we have
$$
D(z_3-z_1)D(z_4-z_2)=\coef
\left[t_1t_2(\frac{y_-}{\n_-}t_3-t_1)(\frac{y_-}{\n_-}t_4-t_2) \right]^{1-D/2}
\,.
$$
This diagram has two cusps singularities for $z_3-z_1\to 0$ and
$z_4-z_2\to 0$ and two light-cone collinear singularities for $z_1\to 0$
and $z_2\to 0.$ Evaluating the integral we obtain
\baa
I_6&=&\Gamma^2(D/2-1)
   \left\{\frac1{(4-D)^4}
    \left(
       \frac{\Gamma(5-D)\Gamma(2D-7)}{2\Gamma(D-3)}
      -\frac{\Gamma(7-3D/2)\Gamma(2D-7)}{3\Gamma(D/2-1)}
\right.\right.
\\
& &\hspace{-10mm}-\left.\left.
          \frac14(2-D)(3-D/2)\Gamma(D-3)
          \left(\Gamma(5-D)-\frac{\Gamma(3-D/2)}{\Gamma(D/2-1)}
          \right)
    \right)
    -\frac1{4-D}\frac{\zeta(3)}2
%   + const.
\right\}
+\CO((4-D)^0)
\,,
\eaa
where $\zeta(3)$ is the Reimann function.

\subsection{Self-energy diagrams}

The general form of the contribution of the diagrams with gluon
self-energy in figs.~6.7 and 6.8 is ($a=7,8$)
$$
W_a=\left(\gdo\right)^2 C_F\left((3D-2)C_A-2(D-2)N_f\right) [2i(\rho -i0)]^{8-2D}
I_a(D) \,.
$$
The one-loop correction to the gluon propagator in the Feynman gauge
in the coordinate representation is given  by
\be
D^{(1)}(z)=\frac{g^2}{64\pi^D}\left((3D-2)C_A-2(D-2)N_f\right)
\frac{\Gamma^2(D/2-1)}{(D-4)(D-3)(D-1)}
(-z^2+i0)^{3-D}
\lab{pro}
\ee
which differs from the free propagator in the power of $z^2-i0$. For
time-like vector $z$ the cutted one-loop propagator coincides with
the full propagator in \re{pro}. As in the one loop case this allows
us to treat the sum of these diagrams with all possible cuts by using
the full propagator. We obtain
$$
W_7 = g^2C_F(\n y)\int_{-\infty}^0 dt_1\int_0^1 dt_2\ D^{(1)}(z_2-z_1)
$$
where $z_1=\n t_1$ and $z_2=y t_2$ and
$$
W_8 = -g^2C_F(\n y)\int_{-\infty}^0 dt_1\int_0^\infty dt_2\
       D^{(1)}_+(z_2-z_1)
$$
where $z_1=\n t_1$ and $z_2=y-\n t_2.$ By performing the integration we get
$$
I_7=\frac{\Gamma^2(D/2-1)\Gamma(5-D)\Gamma(2D-7)}{16(4-D)^3(1-D)\Gamma(D-2)}
\,,
\qqquad
I_8=-\frac{\Gamma^2(D/2-1)\Gamma(5-D)\Gamma(2D-7)}{8(4-D)^2(1-D)\Gamma(D-2)}
$$

\subsection{Diagrams with three-gluon vertices}

For the diagrams of figs.~6.9--6.11 containing three-gluon vertex we have the
following general expression
$$
W_a=\frac12 C_AC_F\left(\gdo\right)^2 [2i(\rho-i0)]^{8-2D} I_a(D)
$$
where $a$ refers to the various diagrams in fig.~6 with three-gluon vertex.
We simplify the analysis of the diagrams with various cuts by computing only
the contributions which are singular for $D\to 4$. In this case we can treat
all gluons as virtual. To show this observe that for the cutted diagrams at
$D=4$ we have only infrared and light-cone collinear singularities. The
infrared singularities cancel. The collinear singularities appear when
gluons propagate along the light-cone, \ie when the intermediate point $z_4$
lies on the segment $[0,y].$ In this case cutted propagators coincide
with the full propagators. Notice that cusp singularities appear when all
gluons interact at small distances. They are present only for diagrams of
figs.~6.9 and 6.11 for $z_i\to 0.$ In this case all gluons are virtual.
Therefore, in the following we study only the contributions to $W_a$ in
which all gluons as virtual.

For the diagram of fig.6.9 we have
$$
W_9=\half g^4 C_AC_F \int_{-\infty}^0 dt_1\int_0^1 dt_2\int_{t_2}^1 dt_3
    \int d^D z_4 \,
    \n^{\mu_1}y^{\mu_2}y^{\mu_3}\Gamma_{\mu_1\mu_2\mu_3}(z_1,z_2,z_3)
    \prod_{i=1}^3 D(z_i-z_4)
$$
where $z_1=\n t_1,$ $z_2=y t_2$ and $z_3=y t_3.$ By using the expression for the
three-gluon vertex in the Appendix~A we find
\be
\n^{\mu_1}y^{\mu_2}y^{\mu_3}\Gamma_{\mu_1\mu_2\mu_3}(z_1,z_2,z_3)
=i(\n y)\left(y\frac{\partial}{\partial z_2}
            -y\frac{\partial}{\partial z_3}
       \right)
=i(\n y)\left(\frac{\partial}{\partial t_2}
            -\frac{\partial}{\partial t_3}
       \right)
\lab{glu1}
\,.
\ee
Due to this particularly simple form of the three-gluon vertex the
integration over $t_2$ or $t_3$ becomes trivial. For the first term the
integration over $t_2$ gives the contributions from the end points $z_2=0$
and $z_2=z_3.$ For the second term the integration over $t_3$ gives the
contributions from $z_3=z_2$ and $z_3=1.$ In all contributions, the
integral over the intermediate point $z_4$ is factorized into the following
expression
\ba
J(z_1,z_2,z_3)&=&\int d^D z_4 \ \ \prod_{i=1}^3 D(z_i-z_4)
\nonumber
\\
& &\hspace{-16mm}= \frac{i^{1-2D}}{32\pi^D}\frac{\Gamma(D-3)}{4-D}
            \int_0^1 ds (s(1-s))^{D/2-2}
            \left((-z_1+sz_2+(1-s)z_3)^2-i0\right)^{3-D}
\lab{id}
\ea
valid for $z_2$ and $z_3$ on the light-cone ($z_2^2=z_3^2=(z_2-z_3)^2=0$). The
pole at $D=4$ corresponds to a light-cone singularity as  $z_4$ approaches the
segment $[0,y].$ The meaning of the integral over the parameter $s$ is the
following. For $D=4$ one integrates over the free gluon propagator between the
points $z_1$ and $z=s z_2+(1-s)z_3$. Since the $z$ lies on the light-cone between
$z_2$ and $z_3$ the vector $z-z_1$ is time-like. This confirms the expectation
expressed at the beginning of this subsection that all collinear gluons propagate
in the time-like direction.

Performing the remaining integration we obtain
$$
I_{9}=-\frac{\Gamma(5-D)\Gamma(2D-7)}{4(4-D)^3}
       \left(\frac{\Gamma^2(D/2-1)}{(4-D)\Gamma(D-3)}
            -\frac{4\Gamma(7-3D/2)\Gamma(D/2-1)}{3(4-D)\Gamma(5-D)}
            +\frac{\Gamma^2(D/2-1)}{\Gamma(D-2)}
       \right)
\,.
$$
For the diagram of fig.~6.10 one deforms the integration path over $z_3=y-\n
t_3$ into $z_3=y+\n t_3$, replaces cutted propagators by full ones and
obtains
$$
W_{10}=\half g^4 C_AC_F \int_{-\infty}^0 dt_1\int_0^1 dt_2\int_0^\infty dt_3
    \int d^Dz_4 \
    \n^{\mu_1}y^{\mu_2}n^{\mu_3}\Gamma_{\mu_1\mu_2\mu_3}(z_1,z_2,z_3)
    \prod_{i=1}^3 D(z_i-z_4)
\,,
$$
where $z_1=\n t_1,$ $z_2=y t_2$ and $z_3=y+\n t_3.$
By using the three-gluon vertex we find
\baa
\n^{\mu_1}y^{\mu_2}n^{\mu_3}\Gamma_{\mu_1\mu_2\mu_3}(z_1,z_2,z_3)
&=&i\left(y\frac{\partial}{\partial z_1}
            -y\frac{\partial}{\partial z_3}
       \right)
 -i(\n y)\left(\n\frac{\partial}{\partial z_1}
            -\n\frac{\partial}{\partial z_3}
       \right)
\\
&=&i\left(y\frac{\partial}{\partial z_1}
            -y\frac{\partial}{\partial z_3}
       \right)
 -i(\n y)\left(\frac{\partial}{\partial t_1}
             -\frac{\partial}{\partial t_3}
       \right).
\eaa
The first term leads to a contribution which is regular for $D=4.$
To see this notice that the singularities arise when the gluons are
propagating along the light-like vector $y.$
However, this configuration is suppressed by applying the operator
$y\frac{\partial}{\partial z_i}$ to the gluon propagator.
The expression $y\frac{\partial}{\partial z_i}D(z_i-z_4)$ is proportional
to $y(z_i-z_4)$ and vanishes for $z_i-z_4$ parallel to $y.$
The second term is similar to the one of the previous diagram.
We have contributions from the end-points $z_1=0$ and $z_3=y.$
For $z_1=0$ ($z_3=y$) the two vectors $z_1$ and $z_2$ ($z_2$ and $z_3$)
lie on the light-cone and we can apply the identity \re{id}.
Performing the remaining integrations we obtain
$$
I_{10}=-\frac{\Gamma(5-D)\Gamma(2D-7)\Gamma(D/2-1)}{2(4-D)^4}
     \left(\frac{\Gamma(D/2-1)}{\Gamma(D-3)}-
           \frac{\Gamma(7-3D/2)}{\Gamma(5-D)}\right).
$$
For the last diagram of fig.~6.11 we have
$$
W_{11}=\half g^4 C_AC_F
       \int_{-\infty}^0 dt_1\int_{t_1}^0 dt_2\int_0^\infty dt_3
       \int d^D z_4\,\n^{\mu_1}\n^{\mu_2}y^{\mu_3}
       \Gamma_{\mu_1\mu_2\mu_3}(z_1,z_2,z_3)
       \prod_{i=1}^3 D(z_i-z_4)\,,
$$
where $z_1=\n t_1,$ $z_2=\n t_2$ and $z_3=yt_3.$ By using three-gluon
vertex we obtain
\ba
\n^{\mu_1}\n^{\mu_2}y^{\mu_3}\Gamma_{\mu_1\mu_2\mu_3}(z_1,z_2,z_3)
&=&-i\left(y\frac{\partial}{\partial z_1}
            -y\frac{\partial}{\partial z_2}
       \right)
 +i(\n y)\left(\n\frac{\partial}{\partial z_1}
            -\n\frac{\partial}{\partial z_2}
       \right)
\nonumber
\\
&=&-i\left(y\frac{\partial}{\partial z_1}
            -y\frac{\partial}{\partial z_2}
       \right)
 +i(\n y)\left(\frac{\partial}{\partial t_1}
            -\frac{\partial}{\partial t_2}
       \right)
\lab{glu2}
\,.
\ea
For this diagram we have the following singularities: a cusp singularity
for $z_i\to 0$ with $i=1,\ldots, 4$; two independent light-cone collinear
singularities for $z_1,z_2\to 0$ and $z_4$ approaching the segment $[0,y]$;
an ultraviolet singularity from $z_2,z_4\to z_1.$

The operator $y\frac{\partial}{\partial z_i}$ in the first term suppresses
propagation along light-like vector $y$ of gluon from point $z_4$ to $z_1$ or to
$z_2.$ This implies that the contribution of this part of three-gluon vertex
contains only a triple pole in $4-D.$ The second term in \re{glu2} is similar to
the one in \re{glu1} for diagram of fig.~6.9. We have two end-point contributions
with $z_2=0$ and $z_1=z_2.$ For $z_2=0$ the two vectors $z_2$ and $z_3$ lie on
the light-cone and we can apply the identity \re{id}. For $z_1=z_2$ the integral
is similar to the one of the gluon self-energy correction.

We finally obtain
\be\label{I11}
I_{11}=\frac{\Gamma(5-D) \Gamma( 2D-6)}{4(4-D)} \left({\frac {\Gamma( 3D/2-5)
\Gamma( 3-D/2) }{3\Gamma(D-2)( 4-D) ^{3}}}-\frac12\,\zeta(3)\right)\,.
\ee
Recall that all expressions in this subsection are valid up to terms
which are regular for $D=4$.

\subsection{Renormalization at two-loop order}

Since the diagrams of fig.~6 have nested ultraviolet divergences corresponding to
the renormalization of the vertices and propagators one has to include additional
counter-terms. In the $\rm \MS-$scheme their contributions are given to two loops
by
$$
w_{c.t.}^{(2)}=\left(\alpi\right)^2 C_F\left(\frac{11}3C_A-\frac23 N_f\right)
[2i(\rho-i0)]^{4-D}\frac{\Gamma(3-D/2)\Gamma(D-3)}{(4-D)^2}
\left(\frac1{4-D}-\frac12\right)\,.
$$
To obtain the final expression for the renormalized $w$ to two-loops we add the
counter-terms, subtract the poles in the $\rm \MS-$scheme and take into account
combinatorial factors. By using the non-abelian exponentiation theorem $w^{(2)}$
is obtained by omitting the colour factor $C_F^2$ in \re{abe}. The final
contributions to $w^{(2)}$ from the various diagrams $(a=1,\ldots,11)$ have the
following form
$$
w_a^{(2)}=\left(\alpi\right)^2C_F\left[
C_A( A_a L^4 + B_a L^3 + C_a L^2 + D_a L )
+N_f( E_a L^3 + F_a L^2 + G_a L )
+ \CO(L^0)\right]
$$
where $L$ is given in \re{res:re} and for the various diagrams
the nonvanishing coefficients are given by
$$
       A_6=-\fracs19,
\qquad A_9=\fracs1{18},
\qquad A_{11}=\fracs1{18},
$$
$$
       B_1=-\fracs29,
\qquad B_7=-\fracs59,
\qquad B_9=-\fracs13,
\qquad B_{11}=-\fracs19,
\qquad B_{c.t.}=\fracs{11}{18},
$$
$$
       C_1=-\fracs13,
\qquad C_2=1,
\qquad C_3=-\fracs{\pi^2}6,
\qquad C_6=-\fracs7{72}\pi^2,
\qquad C_7=-\fracs{31}{36},
$$
$$
       C_8=\fracs56,
\qquad C_9=\fracs{25}{144}\pi^2-\fracs12,
\qquad C_{10}=\fracs1{12}\pi^2,
\qquad C_{11}=\fracs{13}{144}\pi^2-\fracs16,
\qquad C_{c.t}=-\fracs{11}{12},
$$
$$
       D_1=-\fracs{13}{72}\pi^2-\fracs13,
\quad D_3=2\zeta(3),
\quad D_4=\fracs12,
\quad D_6=\fracs29\zeta(3),
\quad D_7=-\fracs{47}{54}-\fracs5{16}\pi^2,
\quad D_8=\fracs{31}{36},
$$
$$
       D_9=\fracs1{72}\zeta(3)-\fracs12-\fracs3{16}\pi^2,
\qquad D_{10}=-\fracs14\zeta(3), \qquad
D_{11}=-\fracs{13}{144}\pi^2+\fracs{19}{72}\zeta(3)-\fracs16, \qquad
D_{c.t.}=\fracs{55}{144}\pi^2,
$$
$$
       E_7=\fracs29,
\qquad E_{c.t.}=-\fracs19,
$$
$$
       F_7=\fracs5{18},
\qquad F_8=-\fracs13,
\qquad F_{c.t.}=\fracs16,
$$
$$
       G_7=\fracs18\pi^2+\fracs7{27},
\qquad G_8=-\fracs5{18},
\qquad G_{c.t.}=-\fracs5{72}\pi^2.
$$
Summing all contributions we finally obtain
\be
w^{(2)}= \left(\alpi\right)^2C_F
\left( B L^3 + C L^2 + D L + \CO(L^0) \right)
\lab{res:main}
\ee
where
\baa
B&=&-\fracs{11}{18}C_A+\fracs19 N_f,
\\
C&=&\left(\fracs1{12}\pi^2-\fracs{17}{18}\right)C_A
       +\fracs19 N_f,
\\
D&=&\left(\fracs94\zeta(3)-\fracs7{18}\pi^2-\fracs{55}{108}\right)C_A
 +\left(\fracs1{18}\pi^2-\fracs1{54}\right) N_f
\,.
\eaa
This expression has the following properties: \bit \item the coefficient of $L^4$
vanishes; \item the coefficient of $L^3$ is proportional to the one-loop
beta-function; \item the interpretation of the remaining coefficients in front of
$L^2$ and $L$ will be clear from the RG equation for light-like Wilson loop we
are going to discuss in sect.~4. \eit

The two-loop calculation confirms the analytical dependence in \re{ana}
which means that all singularities of $W(C_S)$ in the complex $\rho-$plane
lie in the upper half plane. After integration over $y_-$ in \re{main}
this leads to the spectral property
\be
F(x,\mu/M)=0 \qquad \mbox{for $x > 1.$}
\lab{spe}
\ee
It has been observed \ci{break} that the presence in \re{res:main} of the $L^3$
term seems to be in contradiction with the renormalization properties of Wilson
loops. Assuming that $W(C_S)$ is renormalized multiplicatively, the $L^3$ term
should vanish. However one should notice that we are dealing with a light-like
Wilson loop which has additional light-cone singularities. Moreover, the fact
that the coefficient of $L^3$ is just the one-loop beta-function suggests that
light-like Wilson loop obeys a RG equation. This equation is discussed in the
next section.

\sect{Renormalization group equation for Wilson loop on the light-cone}

To evaluate the structure function for $x\to 1$ given in eq.\re{main} one needs
to compute $W(C_S)$ in \re{wl} to all orders in perturbation theory. A powerful
method to resum the expansion is the use of renormalization group equation. In
this section, following ref.~\ci{WL}, we deduce the RG equation of Wilson loops
with path partially lying on the light-cone.

Recall that for $x$ away from 1 the operator product expansion on the
light-cone allows us to relate the $\mu-$dependence of the structure
function with the ultraviolet properties of local composite twist-2
operators obtained by expanding in powers of $y_-$ the matrix element in
\re{str}. This dependence is described by the evolution equations \ci{EQ}
\be
\left(\mu\partder{\mu}+\beta(g)\partder{g}\right)F(x,\mu/M) =\int_x^1 dz\ P(x/z)
F(z,\mu/M)\,, \lab{eq1}
\ee
where the splitting function $P(z)$ is singular for $z\to1$. For $x\to 1$
the structure function \re{main} is given in terms of $W(C_S)$, which
is a nonlocal operator. This suggests that it is convenient to treat
the Wilson loop as a nonlocal functional of gauge field rather than to
expand it into sum of infinitely many local composite operators. This can
be done by using RG equation of $W(C_S)$, which gives the dependence on
the renormalization point $\mu$. Since $W(C_S)$ is a function of the
single parameter $\rho \sim \mu y_-$, from its $\mu$ dependence we
directly obtain the dependence on $y_-$ and, through Fourier transform in
\re{main}, the dependence on $x$ for the structure function.

\subsection{Wilson loop renormalization away from the light-cone}

Renormalization group equation for the Wilson loop away from the light-cone
is well known \ci{away} and depends on the explicit form of the path.
The integration path for $W(C_S)$ is shown in fig.~2. It has two cusps at
the points $0$ and $y$ where the quark is probed by the photon. The
important property of this path is that the segment $\ell_2$ lies on
the light-cone. As we discussed in the two loop calculation, the presence
of the $L^3$ term entails that the renormalization properties of $W(C_S)$ are
different from the ones of Wilson loops with path away from the light cone.

Suppose for a moment that $\ell_2$ lies away from the light-cone, \ie
$y^2\neq 0.$ Then, the dependence on $\mu$ of the renormalized Wilson loop
expectation value is described by the RG equation
\be
\left(\mu\frac{\partial}{\partial\mu} +\beta(g)\frac{\partial}{\partial g}\right)
\ln W_{(y^2\neq 0)}(C_S)=-\Gamma_{\rm cusp}(\gamma_+,g)-\Gamma_{\rm
cusp}(\gamma_-,g)\,, \lab{rg1}
\ee
where $\gamma_\pm$ are the angles in Minkowski space between vectors $\pm y_\mu$
and $\n_\mu=p_\mu/M$ with $\n^2=1$
$$
\cosh \gamma_\pm=\pm\frac{(\n y)}{\sqrt{y^2}}\,.
$$
The cusp anomalous dimension $\Gamma_{\rm cusp}(\gamma,g)$ is gauge invariant
function of the cusp angle \ci{away}. In the limit of large angles $\gamma$ we
have \ci{cusp}
$$
\Gamma_{\rm cusp}(\gamma,g)=\gamma \Gamma_{\rm cusp}(g)+\CO(\gamma^0)\,,
$$
where the coefficient $\Gamma_{\rm cusp}(g)$ is known to two loop order and is
given by
\be
\Gamma_{\rm cusp}(g)=\alpi C_F+\left(\alpi\right)^2
C_F\left(C_A\left(\frac{67}{36}-\frac{\pi^2}{12}\right)-N_f\frac5{18}\right).
\lab{cusp}
\ee
For the integration path of fig.~2 we have $y^2=0$ leading to an infinite
$\gamma_\pm,$ thus eq.\re{rg1} becomes meaningless. One can directly check that
the two-loop result of previous section does not satisfy equation \re{rg1}.
However, as shown in subsect.~4.2, there is a simple way to find the
generalization of the renormalization group equation for the light-cone Wilson
loop.

\subsection{Wilson loop renormalization on the light-cone}

Renormalization group equation for Wilson loops on the light-cone has been
proposed in \ci{WL} and can be obtained as follows. First, one slightly shifts
the integration path away from the light-cone by setting $y^2\neq 0$ and keeping
the cusp angle $\gamma_\pm=\half\ln(4(\n y)^2/y^2)$ large. Since $\gamma_\pm$ is
a logarithmic function of $(\n y)$ and $y^2$ we need to define its analytical
continuation away from positive $(\n y)$ and $y^2.$ The proper expressions for
$\gamma_\pm$ can be deduced from one-loop calculation of Wilson loop \ci{WL} and
are given by
$$
\gamma_+=\half\ln\frac{4((\n y)-i0)^2}{y^2-i0}\,, \qqquad
\gamma_-=\half\ln\frac{4((\n y)+i0)^2}{y^2-i0}
$$
where the ``$-i0$'' prescription comes from the position of the
singularity of the free gluon propagator in the coordinate representation.
By using these expressions we now differentiate the renormalization group
equation \re{rg1} with respect to the variable $(\n y)$
\be
\left(\mu\frac{\partial}{\partial\mu}+\beta(g)\frac{\partial}{\partial g}
\right)\frac{\partial}{\partial (\n y)} \ln W(C_S) =-\Gamma_{\rm
cusp}(g)\left(\frac{1}{(\n y)-i0} +\left(\frac{1}{(\n
y)+i0}\right)^\dagger\right) =-\frac{2\Gamma_{\rm cusp}(g)}{(\n y)-i0} \lab{rg2}
\,.
\ee
The two terms originate from two cusps of the path of fig.~2 which lie
on the opposite sides of the cut. This is the reason for the appearance
of complex conjugation in the second term.

The variable $y^2$ disappeared from this equation and one can formally
set $y^2=0.$ This is the proposed renormalization group equation for
the Wilson loop on the light-cone \ci{WL}. One easily check that two-loop
expression \re{res:main} of the previous section does satisfy eq.\re{rg2}.

Notice, that Wilson loop on the light-cone depends only on a single variable
$L=\ln(i(\rho-i0)) + \gamma_{\rm E}.$ Thus equation \re{rg2} becomes very
powerful since one can integrate it and obtain RG equation for the light-like
Wilson loop
\be
\left(\mu\frac{\partial}{\partial\mu}+\beta(g)\frac{\partial}{\partial g} \right)
W(C_S) =-\left[2\Gamma_{\rm cusp}(g)L + \Gamma(g)\right] W(C_S)\,, \lab{rg3}
\ee
where $\Gamma(g)$ is the integration constant. From this equation we can see the
origin of the various terms in the two-loop calculations in \re{res:main}. The
appearance of the one-loop beta function in front of the $L^3$ term is obvious
from \re{rg3}. The coefficient of $L^2$ is proportional to a sum of $\Gamma_{\rm
cusp}(g)$ and one-loop beta-function. The coefficient of $L$ is given by
\be
\Gamma(g)=-\alpi C_F +\left(\alpi\right)^2C_F\left[
\left(\frac{55}{108}+\frac1{144}\pi^2-\frac94\zeta(3)\right)C_A
 +\left(\frac1{54}+\frac1{72}\pi^2\right) N_f
\right]. \lab{Gamma-new}
\ee
While $\Gamma_{\rm cusp}(g)$ is a universal number, $\Gamma(g)$ depends on the
path under consideration. The unusual feature of the RG equation \re{rg3} is that
the anomalous dimension given by the coefficient of $W(C_S)$ in the r.h.s.
depends on $\rho$, \ie on the renormalization point $\mu$ and $y_-.$ Actually
this property leads to the evolution equation for the structure function as we
shall discuss in the next section.

\sect{Evolution equations}

The RG equation for the distribution function near the phase space boundary is
obtained by using the representation \re{main} and properties \re{res:re'} and
\re{ana}. Introducing the dimensionless parameter $\sigma=(Py)$ we can write
\be
F(x,\mu/M)=H(\mu/M)\int_{-\infty}^{\infty} \frac{d\sigma}{2\pi} \,
\e^{i\sigma(1-x)} W(\sigma\mu/M-i0) \lab{F-Four}
\ee
with the inverse transformation
$$
H(\mu/M)W(\sigma\mu/M-i0)= \int_{-\infty}^1 dx\ \e^{-i\sigma(1-x)}F(x,\mu/M)\,,
$$
where the range of $x-$integration takes into account the spectral
property of the structure function in \re{spe}. The renormalization group
equation in \re{rg3} gives
\be
{\cal D} \ F(x,\mu/M)\equiv\left(\mu\partder{\mu}+\beta(g)\partder{g}\right)
F(x,\mu/M)
=\int_{x}^1 dz\ P(1+x-z) F(z,\mu/M)
\lab{eq2}
\ee
where
$$
P(z)=\int_{-\infty}^{\infty} \frac{d\sigma}{2\pi} \
\e^{i\sigma(1-z)}\left\{-2\Gamma_{\rm cusp}(g)\ln[i(\sigma\mu/M-i0)\e^{\gamma_E}]
-\Gamma(g)+{\cal D} \ \ln H(\mu/M)\right\}\,.
$$
{}  From the analytical property of the integrand we have
$$
P(z)=0 \qquad \mbox{for $z > 1$}\,,
$$
this is the reason for setting to $x$ the lower limit for $z$
in \re{eq2}. To compute $P(z)$ we use the representation
$$
\ln \rho =\int_0^\infty\frac{d\alpha}{\alpha} \left(\e^{- \alpha}-\e^{-
\alpha\rho}\right)\,.
$$
After a careful treatment of the $\alpha \to 0$ singularities, we  obtain
\be
P(z)=2\Gamma_{\rm cusp}(g)\left(\frac{\theta(1-z)}{1-z}\right)_+
     +\delta(1-z) h(g)
\lab{P-res}
\ee
where $(\ldots)_+$ is the standard plus-distribution and
\be
h(g)=-\Gamma(g)+2\Gamma_{\rm cusp}(g)\ln\frac{M}{\mu} +{\cal D}\; \ln H(\mu/M)\,.
\lab{h}
\ee
Due to the factorization theorem \ci{FT}, $P(z)$ should not depend on $\mu$ and
the same is then true for the function $h(g)$. This means that the $\mu$
dependence in the term involving the coefficient function $H(\mu/M)$ should be
compensated by a contribution from $W(C_S)$. One can consider now \re{h} as the
RG equation for the function $H(\mu/M)$. While only soft gluons contribute to
$\Gamma_{\rm cusp}(g)$, both soft and collinear gluons (and quarks) contribute to
$h(g)$. Therefore the function $h(g)$ is not fixed by the RG equation of $W(C_S)$
and should be directly computed \ci{2L} from Feynman diagrams.

Notice that the function $F(x,\mu/M)$ defined in \re{F-Four} satisfies the
evolution equation \re{eq2} for any value of $x$. However, this function has the
physical meaning of the quark distribution only for $x\to 1$.

The evolution equation \re{eq2} in the soft limit $x\sim 1$ and $z\sim 1$ can be
written in the standard form \re{eq1} where $P(z)$ is the two-loop quark
splitting function for $z$ near $1.$ From \re{P-res} the general form of $P(z)$
is given by
\be\lab{pz}
(1-z) P(z)=2\Gamma_{\rm cusp}(g)\; \theta(1-z)\,.
\ee
Using the expression \re{cusp} for $\Gamma_{\rm cusp}(g)$ one easily verifies
that the result of two-loop calculations \ci{2L} obeys this relation.

The evolution equation \re{eq2} for $x\sim 1$ is obtained from the RG
equation for the light-like Wilson loop. Because of the universal
structure of RG equation, one finds that any distribution, which can
be represented in terms of the light-like Wilson loops, satisfies equation
\re{eq2} with the kernel $P(z)$ defined by \re{pz}. This is the case for
the quark and gluon structure and fragmentation functions (see subsect.~2.3)
at large $x$. For the gluon distributions one should replace the colour
factor $C_F$ by $C_A$ in \re{cusp}.

Let us consider the moments of the function $F(x,\mu/M)$ defined in \re{F-Four},
\be
F_n(\mu/M) = \int_0^1 dx\, x^{n-1} F(x,\mu/M) = H(\mu/M) \int_0^1 dx\, x^{n-1}
\int_{-\infty}^\infty \frac{d\sigma}{2\pi}\, \e^{i\sigma(1-x)}
W(\sigma\mu/M-i0)\,. \lab{F-mom}
\ee
For large $n$ the integral over $x$ receives the leading contribution from the
$x\to 1$ region in which $F(x,\mu/M)$ coincides with the quark distribution
function. By using the two-loop expression for $W(\sigma\mu/M-i0)$ in
\re{res:main} we obtain the corresponding $F_n(\mu/M)$. Before this we note that
$W(\sigma\mu/M-i0)$ is given to any order in perturbation theory by a sum of
powers of $L=\ln[i(\sigma\mu/M-i0)]+\gamma_E$. Hence, in order to find
$F_n(\mu/M)$ one needs to expand the following basic integral
$$
\int_0^1 dx\, x^{n-1} \int_{-\infty}^\infty \frac{d\sigma}{2\pi}\,
\e^{i\sigma(1-x)} \left[i\left(\sigma\frac{\mu}{M}-i0
\right)\right]^{-\delta}=\left(\frac{\mu}{M}\right)^{-\delta}\frac{\Gamma(n)}{\Gamma(n+\delta)}
=\left(\frac{\mu}{M}n\right)^{-\delta} \left( 1 + \CO(1/n)\right),
$$
in powers of $\delta$. This relation implies that for large $n$ and arbitrary
$\delta$ the integral is given by the integrand evaluated at $\sigma=-in$.
Performing this trick in \re{F-mom} we get the simple expression
\be
F_n(\mu/M) = H(\mu/M) W(-in\mu/M)\left( 1 + \CO(1/n)\right), \lab{F=W}
\ee
which means that the large$-n$ behavior of the moments of the quark distribution
function is given by the Wilson loop expectation value evaluated along the path
of fig.~2 with formal identification $(Py) = -in$ (or $\rho=-in\mu/M$) in
\re{res:re'} and \re{ana}. From \re{F=W} and the non-abelian exponentiation
theorem we obtain that large$-n$ corrections to $F_n$ exponentiate. From the one-
and two-loop results in \re{res:re} and \re{res:main} we get that the exponent of
$F_n$ contains a sum of powers of $\ln n$ up to $\as\ln^2 n$ terms in one loop
and $\as^2\ln^3 n$ terms in two loops. The sum of all these corrections to all
orders can be done by using the RG equation \re{rg3} for light-like Wilson loops.
From this equation and from \re{F=W} we find that the large$-n$ behaviour of
$F_n(\mu/M)$ is governed by
\be
\left(n\frac{\partial}{\partial n} +\beta(g)\frac{\partial}{\partial g}  \right)
F_n(\mu/M) = -\left[2\Gamma_{\rm cusp}(g)\ln(n\e^{\gamma_{\rm E}}\mu/M)+\Gamma(g)
\right]F_n(\mu/M)\,. \lab{EQ1}
\ee
To test the relation \re{F=W} we differentiate $F_n(\mu/M)$ w.r.t.\ $\mu$ and
substitute eqs.~\re{rg3} and \re{h} to obtain
\be
\left(\mu\frac{\partial}{\partial \mu} +\beta(g)\frac{\partial}{\partial g}
\right) F_n(\mu/M) = -\left[2\Gamma_{\rm cusp}(g)\ln(n\e^{\gamma_{\rm E}})-h(g)
\right]F_n(\mu/M)\,.\lab{EQ2}
\ee
One easily checks that the same equation follows from \re{eq2} with the l.h.s.\
equal to the moment of the evolution kernel \re{P-res} for large $n$.

\sect{Concluding remarks}

To conclude we would like to mention a close relation between the analysis here
presented and the heavy quark effective field theory (for a review see
ref.~\ci{Geo}) which seems to be a powerful tool for analyzing of heavy meson
phenomenology. The emission of gluons from heavy quarks can be treated by the
eikonal approximation. This implies that the propagation of the heavy quarks
through the cloud of light particles can be described by Wilson lines and that
the effective heavy quark field theory can actually be formulated \ci{KR} in
terms of the Wilson lines. Therefore, the renormalization properties of Wilson
lines discussed in this paper are related to the ones for the effective theory.
For example one finds that the ``velocity dependent anomalous dimension'' is the
cusp anomalous dimension \re{cusp}.

The central point of our analysis was the RG equation \re{rg3} for the
generalized Wilson loop expectation value, $W(C)$, with path partially lying on
the light-cone. The cusp anomalous dimension $\Gamma_{\rm cusp}(g)$ entering into
this equation is a new universal quantity of perturbative QCD which controls the
behaviour near the phase space boundary of hard distributions. The evolution
equation for these distributions corresponds to the RG group equation for $W(C)$,
moreover the splitting function near the phase space boundary is related to the
cusp anomalous dimension. This equation does not imply that $W(C)$ is
renormalized multiplicatively.

%In particular, at two loop order one finds a non vanishing $L^3$ term with a
%coefficient given by the one-loop $\beta-$function. This follows directly from
%the RG equation of $W(C)$. Moreover $\Gamma_{\rm cusp}(g)$ is contained in the
%coefficient of $L^2$ (see term $C$ in eq.~\re{res:main}).

It is well known that the structure function of deep inelastic scattering gets
large perturbative corrections for $x\to 1$ which need to be
summed~\ci{add2,add3}. They come from two subprocesses: from quark distribution
function and from cross section of the partonic subprocess. In the present paper
we considered the $x\to 1$ behavior of the distribution functions. We were able
to control in eqs.~\re{EQ1} and \re{EQ2} their large perturbative corrections to
all orders using the renormalization properties of light-like Wilson loops. The
same method can be applied for the summation of large perturbative corrections to
the partonic cross section~\ci{add2,add3}. It will be described in the
forthcoming paper.

\section*{Note added on February 4, 2005}

Recently the two-loop calculation of the Wilson loop $W(C_S)$ has been performed
by E.~Gardi in ref.~\cite{EG}. The obtained results for the Feynman integrals
$I_1,\ldots,I_{10}$ agree with our expressions while $I_{11}$ differs by $\sim
1/(D-4)$ term. We repeated the calculation of $I_{11}$ and reproduced the result
of \cite{EG}. In the present version, we updated expressions for the integral
$I_{11}$, Eq.~\re{I11}, the $D-$coefficient, Eq.~\re{res:main} and the anomalous
dimension $\Gamma(g)$, Eq.~\re{Gamma-new}, by taking into account the
contribution of the additional term.

We are most grateful to E.~Gardi and M.~Neubert for useful discussions and
correspondence.

\bigskip\bigskip

\noindent{\Large{\bf Appendix~A:\ \ Feynman rules in the coordinate
representation}}

\bigskip

\setcounter{equation}0

\renewcommand{\theequation}{A.\arabic{equation}}

\noindent In this appendix we recall the Feynman rules for
calculation the generalized Wilson loop expectation value
in the coordinate representation using dimensional regularization.
The $D-$dimensional free gluon propagator in the Feynman gauge
is given by
$D^{\mu\nu}(x)=-g^{\mu\nu}D(x)$ where
$$
D(x)=i\int \frac{d^Dk}{(2\pi)^D}\e^{-ikx}
             \frac{1}{k^2+i0}
= \frac{\Gamma(D/2-1)}{4\pi^{D/2}}(-x^2+i0)^{1-D/2}
$$
The ``cutted'' propagator $D_+^{\mu\nu}(x)=-g^{\mu\nu}D_+(x)$
associated to a real gluon is defined as
$$
D_+(x)=\int \frac{d^Dk}{(2\pi)^D}\e^{-ikx}2\pi\theta(k_0)\delta(k^2)
= \frac{\Gamma(D/2-1)}{4\pi^{D/2}}[-2(x_+-i0)(x_--i0)]^{1-D/2}
$$
where the last equality holds only for $\bom{x}_T=0.$
We note that for gluon propagating in the time-like
direction ($x_+>0,$ $x_->0$ and $\bom{x}_T=0$) cutted and full propagator
coincide. For the three-gluon vertex with three gluon propagators
attached we have

\bigskip
$
\Gamma_{\mu_1\mu_2\mu_3}(z_1,z_2,z_3)\int d^D z_4\
\prod_{i=1}^3 D(z_4-z_i)
$ \hfill

\bigskip
\hfill $=-i\left(
  g^{\mu_1\mu_2}\left(\partial_1^{\mu_3}-\partial_2^{\mu_3}\right)
 +g^{\mu_2\mu_3}\left(\partial_2^{\mu_1}-\partial_3^{\mu_1}\right)
 +g^{\mu_1\mu_3}\left(\partial_3^{\mu_2}-\partial_1^{\mu_2}\right)
  \right) \int d^D z_4 \ \ \prod_{i=1}^3 D(z_4-z_i)
$

\bigskip
\noindent For the gluon attached to the point on the ray $\ell_1$ of the
integration path of fig.~2 and propagating to $z$ we have $ ig\n_\mu
\int_{-\infty}^0 dt_1 D(\n t_1 -z) \,. $ The analogous expressions for a gluon
attached to the segment $\ell_2$ and the ray $\ell_3$ are $ igy_\mu \int_0^1 dt_2
D(y t_2 -z) $ and $ -ig\n_\mu \int_0^\infty dt_3 D(y-\n t_3 -z) $ respectively.

The scalar products and integration measure in terms of light-cone
variables are
$$
a_\pm=\frac1{\sqrt{2}}(a_0\pm a_3),
\qquad
\bom{a}=(a_1,a_2),
\qquad
(a b)= a_+b_-+a_+b_- -\bom{a}\cdot\bom{b},
\qquad
d^Da=da_+\ da_-\ d^{D-2}\bom{a}
$$
for arbitrary $D-$dimensional vectors $a_\mu$ and $b_\mu$.

%\newpage

\newcommand{\bb}{\begin{thebibliography}}
\newcommand{\eb}{\end{thebibliography}}

\bb{99}
\bi{IR} A.H.Mueller, \prep{73}{237}{81}; \\
        G.Altarelli, \prep{81}{1}{82};   \\
        A.\ Bassetto, M.\ Ciafaloni and G.\ Marchesini,
        \prep{100}{201}{83};\\
        Yu.L.\ Dokshitzer, V.A.\ Khoze, A.H.\ Mueller and S.I.\ Troyan,
        \rmp{60}{373}{88};  {\em Basics of Perturbative QCD}\/, Editions Fronti\`eres,
        Paris, 1991.
\bi{add2}
        G.Sterman, Nucl. Phys. B281 (1987) 310.
\bi{add3}
        S.Catani and L.Trentadue, Nucl. Phys. B327 (1989) 323; B353 (1991) 183.
\bi{add4}
        G.Parisi, Phys. Lett. B90 (1980) 295; \\
        G.Curci and M.Greco, Phys. Lett. B92 (1980) 175.
\bi{add5}
        W.L. van Neerven, Phys. Lett. B147 (1984) 175.
\bi{add6}
        S.Catani and M.Ciafaloni, Nucl. Phys. B236 (1984) 61; B249 (1985) 301; \\
        S.Catani, M.Ciafaloni and G.Marchesini, Nucl. Phys. B264 (1986) 558.
\bi{add7}
        S.V.Ivanov and G.P.Korchemsky, Phys. Lett. B154 (1985) 197; {\it in}
        Proc. Quarks 84, Tbilisi, 1984, Vol.2, p.145.

 \bi{EQ}   V.N.Gribov and L.N.Lipatov, Yad. Fiz. 15 (1972) 781; 1218;
\\        L.N.Lipatov, Yad. Fiz. 20 (1974) 181;
\\        G.Altarelli and G.Parisi, Nucl.Phys. 126B (1977) 298.
\bi{FT} J.C.Collins, D.E.Soper and G.Sterman,
        ``Factorization of Hard Processes in QCD,''
        in ``Perturbative Quantum Chromodynamics'', ed. by A.H.Mueller (World
        Scientific, Singapore, 1989) p.1.
\bi{Col} J.C.Collins and D.E.Soper, Nucl. Phys. B194 (1982) 445.
\bi{LNK} T.Kinoshita, J.Math.Phys. 3 (1962) 650;
\\       T.D.Lee and M.Nauenberg, Phys.Rev. 133 (1964) 1549.

\bi{add12}
        G.Sterman, Phys. Rev. D17 (1977) 2773.
\bi{away} A.M.Polyakov, Nucl. Phys. B164 (1980) 171;
\\        I.Ya.Aref'eva, Phys. Lett. B93 (1980) 347;
\\        V.S.Dotsenko and S.N.Vergeles, Nucl. Phys. B169 (1980) 527;
\\        R.A.Brandt, F.Neri and M.-A.Sato, Phys. Rev. D24 (1981) 879.
\bi{ET} J.G.M.Gatheral, Phys. Lett. 113B (1984) 90;
\\      J.Frenkel and J.C.Taylor, Nucl. Phys. B246 (1984) 231.
\bi{break} A.Andrasi and J.C.Taylor, Nucl. Phys. B350 (1991) 73.
\bi{WL} I.A.Korchemskaya and G.P.Korchemsky,
        Phys. Lett. 287B (1992) 169.

\bi{cusp} G.P.Korchemsky and A.V.Radyushkin, Nucl. Phys. B283 (1987)
          342.
\bi{2L}
         G. Curci, W. Furmanski and R. Petronzio, \np{175}{27}{80};
\\       W. Furmanski and R. Petronzio, \zp{11}{293}{82};
\\       J. Kalinowski, K. Konishi, P.N. Scharbach and T.R. Taylor,
       \np{181}{253}{81};
\\       E.G. Floratos, C. Kounnas and R. Lacaze, \pl{98}{89}{81};
\\       I. Antoniadis and E.G. Floratos, \np{191}{217}{81}.
\bi{Geo} For a review see: H.Georgi, ``Heavy Quark Effective
         Field Theory,'' preprint HUTP--91--A039 (1991).
\bi{KR}  G.P.Korchemsky and A.V.Radyushkin, \pl{279}{359}{92}.

\bi{EG}  E.Gardi, %``On the quark distribution in an on-shell heavy quark and its all-order
%relations with the perturbative fragmentation function,''
arXiv:hep-ph/0501257.
%%CITATION = HEP-PH 0501257;%%

\eb

\end{document}